\documentclass[]{article}

\usepackage[margin=1.4in]{geometry}
\usepackage[utf8]{inputenc}
\usepackage[english]{babel}
\usepackage{parskip}
\usepackage{siunitx}

\usepackage{amsmath}
\usepackage{amsfonts}
\usepackage{amsthm}

\usepackage[ruled,vlined,linesnumbered]{algorithm2e}

\usepackage{graphicx}
\usepackage{subcaption}
\usepackage{pgfplots}
\usepackage{pgfplotstable}
\pgfplotsset{width=10cm,compat=1.9}
\usetikzlibrary{patterns}
\usetikzlibrary{graphs}
\pgfplotsset{compat = 1.13,
        my ybar legend/.style={
            legend image code/.code={
                \draw [##1] (0cm,-0.6ex) rectangle +(2em,1.5ex);
            },
        },
}

\usepackage{algorithmic}

\usepackage{import}
\usepackage{cite}
\usepackage{todonotes}
\usepackage{booktabs}
\usepackage{listings}

\title{Parallel two-stage reduction to Hessenberg-triangular form}
\author{Thijs Steel \and Raf Vandebril}

\begin{document}

\maketitle

\begin{abstract}
	We present a two-stage algorithm for the parallel reduction of a pencil to Hessenberg-triangular form. Traditionally, two-stage Hessenberg-triangular reduction algorithms achieve high performance in the first stage, but struggle to achieve high performance in the second stage. Our algorithm extends techniques described by Karlsson et al.\ \cite{karlsson2011parallel} to also achieve high performance in the second stage. Experiments in a shared memory environment demonstrate that the algorithm can outperform state-of-the-art implementations.
\end{abstract}

\section{Introduction}

We consider the parallel reduction of a matrix pencil to Hessenberg-triangular form. For a pencil $(A,B) \, , \, A,B \in \mathbb{C}^{n\times n}$, this consists of finding unitary matrices $Q$ and $Z$, a Hessenberg matrix $H$ and an upper triangular matrix $T$ so that
\begin{equation*}
	Q(H,T)Z^* = (A,B).
\end{equation*}
The most common use for such a decomposition is as a preprocessing step for the QZ algorithm, which solves generalized eigenvalue problems and was introduced by C.B. Moler and G.W. Stewart \cite{molerandstewart}. The goal of this paper is to introduce and evaluate a new Hessenberg-triangular reduction algorithm that scales well on multicore machines.

A common way to parallelize dense linear algebra algorithms is to formulate them so that as many of the flops as possible are within large matrix-matrix multiplications. Implementations of these algorithms can then rely on highly optimized implementations of matrix-matrix multiplications that perform well in parallel. In 2008, B. K\r{a}gstr\"{o}m , D. Kressner, E.S. Quintana-Ort{\'\i}, and G. Quintana-Orti \cite{kaagstrom2008blocked} introduced an algorithm for the reduction of a pencil to Hessenberg-triangular form. Their algorithm performs at least 60\% of its operations via matrix-matrix multiplications. This is a significant improvement over the original reduction by Moler and Stewart, but it is still problematic when we consider parallelization. If we rely only on the parallelization of the matrix-matrix multiplications, then 40\% of the work will not be parallelized.

A possible solution is to use two-stage algorithms. In the first stage, the pencil is reduced to r-Hessenberg-triangular form. A pencil $(H, T)$ is in r-Hessenberg-triangular form if $T$ is upper triangular and $H_{i,j} = 0 \, \forall i > j + r$, i.e. $H$ has at most $r$ nonzero subdiagonals. Efficiently parallelizing a reduction to r-Hessenberg-triangular form is much easier. A parallel algorithm for this reduction was first introduced by K. Dackland and B. K\r{a}gstr\"{o}m  \cite{dackland1998scalapack} and improved upon by K\r{a}gstr\"{o}m  et al.\ \cite{kaagstrom2008blocked}. In the second stage, the pencil in r-Hessenberg-triangular form is then reduced to Hessenberg-triangular form.
To the extent of our knowledge, no effective parallelizations of the second stage are present in the literature. Additionally, the combination of the two stages requires approximately 50\% more flops than the one-stage algorithms. Despite these disadvantages, K\r{a}gstr\"{o}m  et al.\ report that the two-stage approach can achieve good performance.

Similar two-stage algorithms exist for the reduction of a matrix to Hessenberg form. In particular, L. Karlsson and B. K\r{a}gstr\"{o}m \cite{karlsson2010efficient,karlsson2011parallel} were able to efficiently parallelize the second stage for Hessenberg matrices. The main contribution of this paper is to extend their work to Hessenberg-triangular reductions.

Another solution can be found in the work of B. Adlerborn, L. Karlsson, and B. K\r{a}gstr\"{o}m  \cite{adlerborn2018distributed}. They also realized that relying on the parallelization of matrix-matrix multiplications is not an effective strategy. Instead of switching to a two-stage algorithm, they parallelize each step of the one-stage reduction of K\r{a}gstr\"{o}m  et al.\ \cite{kaagstrom2008blocked} with great care.

The rest of this paper is organized as follows. First, we recall the blocked reduction to r-Hessenberg-triangular form by K\r{a}gstr\"{o}m  et al.\ \cite{kaagstrom2008blocked} and discuss ways to parallelize it in Section~\ref{sec: r-hessenberg-triangular}. Second, we introduce a new parallel algorithm to reduce r-Hessenberg-triangular pencils to Hessenberg-triangular form in Section~\ref{sec: hessenberg-triangular}. Finally, we evaluate the performance of the parallel two-stage algorithm in Section~\ref{sec: experiments}.

\section{Reduction to r-Hessenberg-triangular form}
\label{sec: r-hessenberg-triangular}

\subsection{WY form of Householder reflectors}
\label{subsec: wy}

In the algorithms we will present later, we will often need to apply a sequence of Householder reflectors to a matrix. Applying these reflectors one at a time is usually not optimal. For better cache performance, we can use the WY representation of this sequence. If $Q = Q_r Q_{r-1} \dots Q_1$ is the product of these reflectors, then we can factorize $Q$ as
\begin{equation*}
	Q = I - WY^*,
\end{equation*}
where $Q$ is an $n \times n$ matrix and $W$ and $Y$ are $n \times r$ matrices. An algorithm to compute the matrices $W$ and $Y$ is due to C. Bischof and C. Van Loan \cite{bischof1987wy}. Multiplying $Q$ with an $n \times m$ matrix $A$ takes roughly the same amount of flops as applying the reflectors individually, but it can be performed using just two matrix-matrix multiplications instead of $r$ applications of a reflector. If a highly optimized matrix-matrix multiplication is available, using the WY representation is typically much more efficient.

In the rest of this text, when we refer to something as a block reflector, we mean that it is a sequence of reflectors that is represented using its WY representation. 

\subsection{Blocked stage one}
\label{subsec: blockedstage1}

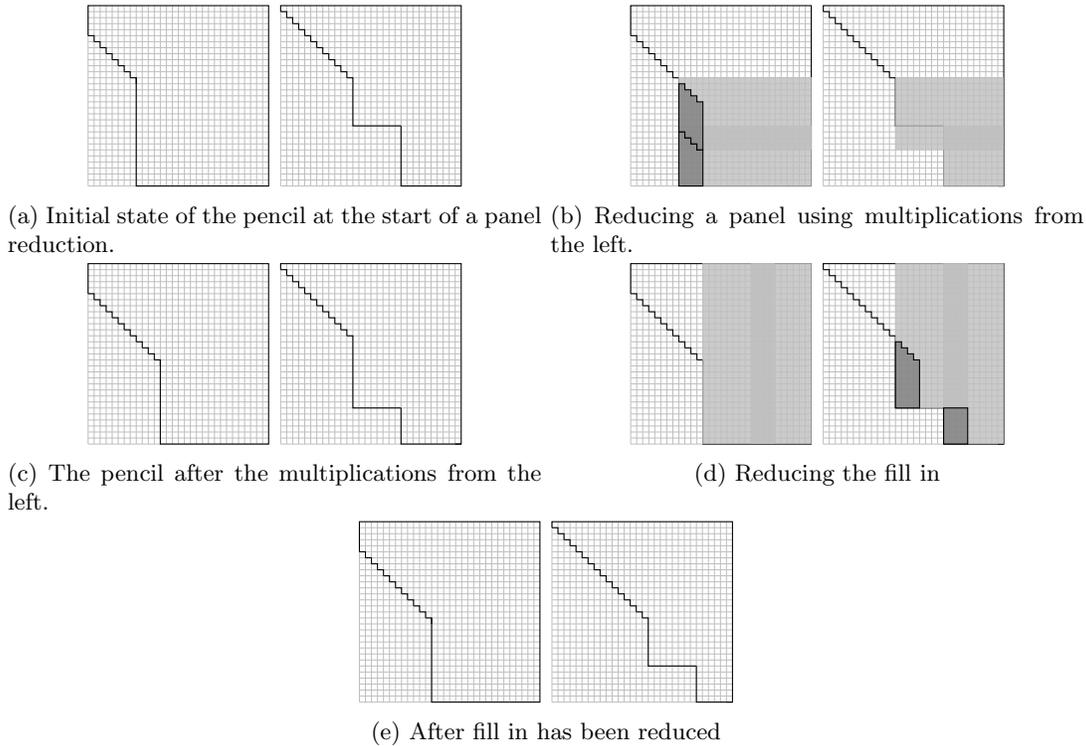
\begin{figure}[htb]
	\centering
		\subcaptionbox{Initial state of the pencil at the start of a panel reduction.\label{subfig:blockedstage1initial}}[.49\textwidth]{%
		
		\begin{tikzpicture}[scale=0.8]
			\foreach \i in {0,1,2,...,29}
				{
					\draw[color=lightgray,line width=0.1mm] (0.1*\i,0) -- (0.1*\i,3);
					\draw[color=lightgray,line width=0.1mm] (0,0.1*\i) -- (3,0.1*\i);
				}

			\draw (0,2.5) -- (0,3) -- (3,3) -- (3,0) -- (2.5,0);
			\foreach \i in {1,2,...,8}
				{
					\draw (0.1*\i,2.5-0.1*\i) -- (0.1*\i,2.5-0.1*\i+0.1) -- (0.1*\i-0.1,2.5-0.1*\i+0.1);
				}
			\draw (2.5,0) -- (0.8,0) -- (0.8,1.8);

			\begin{scope}[shift={(3.2,0)}]
				\foreach \i in {0,1,2,...,29}
					{
						\draw[color=lightgray,line width=0.1mm] (0.1*\i,0) -- (0.1*\i,3);
						\draw[color=lightgray,line width=0.1mm] (0,0.1*\i) -- (3,0.1*\i);
					}
	
				\draw (0,2.9) -- (0,3) -- (3,3) -- (3,0) -- (2.9,0);
				\foreach \i in {1,2,...,12}
					{
						\draw (0.1*\i,2.9-0.1*\i) -- (0.1*\i,2.9-0.1*\i+0.1) -- (0.1*\i-0.1,2.9-0.1*\i+0.1);
					}
				\draw (1.2,1.7) -- (1.2,1.0) -- (2.0,1.0);
				\draw (2.0,1.0) -- (2.0,0) -- (3,0);

			\end{scope}
		\end{tikzpicture}
	}
	\subcaptionbox{Reducing a panel using multiplications from the left.\label{subfig:blockedstage1leftmul}}[.49\textwidth]{%
		
		\begin{tikzpicture}[scale=0.8]
			\foreach \i in {0,1,2,...,29}
				{
					\draw[color=lightgray,line width=0.1mm] (0.1*\i,0) -- (0.1*\i,3);
					\draw[color=lightgray,line width=0.1mm] (0,0.1*\i) -- (3,0.1*\i);
				}

			\draw (0,2.5) -- (0,3) -- (3,3) -- (3,0) -- (2.5,0);
			\foreach \i in {1,2,...,8}
				{
					\draw (0.1*\i,2.5-0.1*\i) -- (0.1*\i,2.5-0.1*\i+0.1) -- (0.1*\i-0.1,2.5-0.1*\i+0.1);
				}
			\draw (2.5,0) -- (0.8,0) -- (0.8,1.8);

			\fill[color=lightgray,opacity=0.8] (0.8,1.0) -- (0.8,0) -- (3,0) -- (3,1.0) -- (0.8,1.0);
			\fill[color=lightgray,opacity=0.8] (0.8,1.8) -- (0.8,0.6) -- (3,0.6) -- (3,1.8) -- (0.8,1.8);

			\draw[fill=gray,fill opacity=0.8] (0.8,1.0) -- (0.8,0) -- (1.2,0) -- (1.2,0.6) -- (1.1,0.6) -- (1.1,0.7) -- (1.0,0.7) -- (1.0,0.8) -- (0.9,0.8) -- (0.9,0.9) -- (0.8,0.9);
			\draw[fill=gray,fill opacity=0.8] (0.8,1.0) -- (0.8,1.7) -- (0.9,1.7) -- (0.9,1.6) -- (1.0,1.6) -- (1.0,1.5) -- (1.1,1.5) -- (1.1,1.4) -- (1.2,1.4) -- (1.2,0.6) -- (1.1,0.6) -- (1.1,0.7) -- (1.0,0.7) -- (1.0,0.8) -- (0.9,0.8) -- (0.9,0.9) -- (0.8,0.9) -- (0.8,1.0);

			\begin{scope}[shift={(3.2,0)}]
				\foreach \i in {0,1,2,...,29}
					{
						\draw[color=lightgray,line width=0.1mm] (0.1*\i,0) -- (0.1*\i,3);
						\draw[color=lightgray,line width=0.1mm] (0,0.1*\i) -- (3,0.1*\i);
					}
	
				\draw (0,2.9) -- (0,3) -- (3,3) -- (3,0) -- (2.9,0);
				\foreach \i in {1,2,...,12}
					{
						\draw (0.1*\i,2.9-0.1*\i) -- (0.1*\i,2.9-0.1*\i+0.1) -- (0.1*\i-0.1,2.9-0.1*\i+0.1);
					}
				\draw (1.2,1.7) -- (1.2,1.0) -- (2.0,1.0);
				\draw (2.0,1.0) -- (2.0,0) -- (3,0);

				\fill[color=lightgray,opacity=0.8] (2.0,1.0) -- (2.0,0) -- (3,0) -- (3,1.0) -- (2.0,1.0);
				\fill[color=lightgray,opacity=0.8] (1.2,1.8) -- (1.2,0.6) -- (3,0.6) -- (3,1.8) -- (1.2,1.8);
			\end{scope}
		\end{tikzpicture}
	}
	\hfill
	\subcaptionbox{The pencil after the multiplications from the left.\label{subfig:blockedstage1afterleftmul}}[.49\textwidth]{%
		
		\begin{tikzpicture}[scale=0.8]
			\foreach \i in {0,1,2,...,29}
				{
					\draw[color=lightgray,line width=0.1mm] (0.1*\i,0) -- (0.1*\i,3);
					\draw[color=lightgray,line width=0.1mm] (0,0.1*\i) -- (3,0.1*\i);
				}

			\draw (0,2.5) -- (0,3) -- (3,3) -- (3,0) -- (2.5,0);
			\foreach \i in {1,2,...,12}
				{
					\draw (0.1*\i,2.5-0.1*\i) -- (0.1*\i,2.5-0.1*\i+0.1) -- (0.1*\i-0.1,2.5-0.1*\i+0.1);
				}
			\draw (2.5,0) -- (1.2,0) -- (1.2,1.4);

			\begin{scope}[shift={(3.2,0)}]
				\foreach \i in {0,1,2,...,29}
					{
						\draw[color=lightgray,line width=0.1mm] (0.1*\i,0) -- (0.1*\i,3);
						\draw[color=lightgray,line width=0.1mm] (0,0.1*\i) -- (3,0.1*\i);
					}

				\draw (0,2.9) -- (0,3) -- (3,3) -- (3,0) -- (2.9,0);
				\foreach \i in {1,2,...,12}
					{
						\draw (0.1*\i,2.9-0.1*\i) -- (0.1*\i,2.9-0.1*\i+0.1) -- (0.1*\i-0.1,2.9-0.1*\i+0.1);
					}
				\draw (1.2,1.7) -- (1.2,0.6) -- (2.0,0.6);
				\draw (2.0,0.6) -- (2.0,0) -- (3,0);

			\end{scope}
		\end{tikzpicture}
	}
	\subcaptionbox{Reducing the fill in\label{subfig:blockedstage1rightmul}}[.49\textwidth]{%
		
		\begin{tikzpicture}[scale=0.8]
			\foreach \i in {0,1,2,...,29}
				{
					\draw[color=lightgray,line width=0.1mm] (0.1*\i,0) -- (0.1*\i,3);
					\draw[color=lightgray,line width=0.1mm] (0,0.1*\i) -- (3,0.1*\i);
				}

			\draw (0,2.5) -- (0,3) -- (3,3) -- (3,0) -- (2.5,0);
			\foreach \i in {1,2,...,12}
				{
					\draw (0.1*\i,2.5-0.1*\i) -- (0.1*\i,2.5-0.1*\i+0.1) -- (0.1*\i-0.1,2.5-0.1*\i+0.1);
				}
			\draw (2.5,0) -- (1.2,0) -- (1.2,1.4);

				\fill[color=lightgray,opacity=0.8] (2.0,0.0) -- (3,0) -- (3,3) -- (2.0,3) -- (2.0,0);
				\fill[color=lightgray,opacity=0.8] (1.2,0) -- (2.4,0) -- (2.4,3) -- (1.2,3) -- (1.2,0);

			\begin{scope}[shift={(3.2,0)}]
				\foreach \i in {0,1,2,...,29}
					{
						\draw[color=lightgray,line width=0.1mm] (0.1*\i,0) -- (0.1*\i,3);
						\draw[color=lightgray,line width=0.1mm] (0,0.1*\i) -- (3,0.1*\i);
					}

				\draw (0,2.9) -- (0,3) -- (3,3) -- (3,0) -- (2.9,0);
				\foreach \i in {1,2,...,12}
					{
						\draw (0.1*\i,2.9-0.1*\i) -- (0.1*\i,2.9-0.1*\i+0.1) -- (0.1*\i-0.1,2.9-0.1*\i+0.1);
					}
				\draw (1.2,1.7) -- (1.2,0.6) -- (2.0,0.6);
				\draw (2.0,0.6) -- (2.0,0) -- (3,0);

				\fill[color=lightgray,opacity=0.8] (2.0,0.0) -- (3,0) -- (3,3) -- (2.0,3) -- (2.0,0);
				\fill[color=lightgray,opacity=0.8] (1.2,0.6) -- (2.4,0.6) -- (2.4,3) -- (1.2,3) -- (1.2,0.6);

				\draw[fill=gray,fill opacity=0.8] (2.0,0.0) -- (2.4,0) -- (2.4,0.6) -- (2.0,0.6) -- (2.0,0);

				\draw[fill=gray,fill opacity=0.8] (1.2,0.6) -- (1.2,1.7) -- (1.3,1.7) -- (1.3,1.6) -- (1.4,1.6) -- (1.4,1.5) -- (1.5,1.5) -- (1.5,1.4) -- (1.6,1.4) -- (1.6,0.6) -- (1.2,0.6);

			\end{scope}
		\end{tikzpicture}
	}
	\subcaptionbox{After fill in has been reduced\label{subfig:blockedstage1afterrightmul}}[.49\textwidth]{%
		
		\begin{tikzpicture}[scale=0.8]
			\foreach \i in {0,1,2,...,29}
				{
					\draw[color=lightgray,line width=0.1mm] (0.1*\i,0) -- (0.1*\i,3);
					\draw[color=lightgray,line width=0.1mm] (0,0.1*\i) -- (3,0.1*\i);
				}

			\draw (0,2.5) -- (0,3) -- (3,3) -- (3,0) -- (2.5,0);
			\foreach \i in {1,2,...,12}
				{
					\draw (0.1*\i,2.5-0.1*\i) -- (0.1*\i,2.5-0.1*\i+0.1) -- (0.1*\i-0.1,2.5-0.1*\i+0.1);
				}
			\draw (2.5,0) -- (1.2,0) -- (1.2,1.4);

			\begin{scope}[shift={(3.2,0)}]
				\foreach \i in {0,1,2,...,29}
					{
						\draw[color=lightgray,line width=0.1mm] (0.1*\i,0) -- (0.1*\i,3);
						\draw[color=lightgray,line width=0.1mm] (0,0.1*\i) -- (3,0.1*\i);
					}

				\draw (0,2.9) -- (0,3) -- (3,3) -- (3,0) -- (2.9,0);
				\foreach \i in {1,2,...,16}
					{
						\draw (0.1*\i,2.9-0.1*\i) -- (0.1*\i,2.9-0.1*\i+0.1) -- (0.1*\i-0.1,2.9-0.1*\i+0.1);
					}
				\draw (1.6,1.3) -- (1.6,0.6) -- (2.4,0.6);
				\draw (2.4,0.6) -- (2.4,0) -- (3,0);



			\end{scope}
		\end{tikzpicture}
	}
	\caption[Illustration of Algorithm \ref{alg:blockedstage1}]{Illustration of the reduction of one panel in stage 1 (Algorithm~\ref{alg:blockedstage1}), with parameters: $n=30$, $n_b=4$, $p=3$. The blocks indicated in gray will be reduced. The updates will affect the parts of the pencil indicated in light gray (the slightly darker gray shows where these updates overlap).}
	\label{fig: blockedstage1}
\end{figure}

Essentially, the blocked stage one algorithm (reduction to r-Hessenberg-triangular form) is a panel reduction. In each iteration, the algorithm reduces a panel of $n_b$ columns of $A$ using block reflectors from the left while using block reflectors from the right to preserve the structure of $B$. The key factor that makes this algorithm a panel reduction is that only submatrices of $A$ and $B$ are updated while calculating the reflectors. After these have been calculated, the rest of the updates can be applied efficiently using matrix-matrix multiplications. Let us consider one iteration of the algorithm, where we want to reduce columns $j$ through $j+n_b-1$. Figure \ref{fig: blockedstage1} illustrates this iteration. The rest of this subsection explains the details of the algorithm.

The initial state of the pencil is shown in Figure~\ref{subfig:blockedstage1initial}. The simplest way to calculate the reflectors from the left would be to take a QR factorization of the block $A(j+n_b:n,j:j+n_b-1)$. The orthogonal factor of this QR factorization reduces an entire panel in $A$. However, applying this block reflector would create a large amount of fill-in in $B$. To avoid this fill-in, we split $A(j+n_b:n,j:j+n_b-1)$ into several blocks of size $pn_b \times n_b$ and take the QR factorization of those blocks, with $p$ an implementation parameter. The reflectors can be calculated while only updating the panel. Afterward, the reflectors can be accumulated and applied to the rest of the pencil using matrix-matrix multiplications. The reduced blocks are illustrated in Figure~\ref{subfig:blockedstage1leftmul}. If $p$ is larger, then the block reflectors are larger and fewer in number, which is typically more efficient. However, if $p$ is large, it will also lead to a large amount of fill-in. K\r{a}gstr\"{o}m et al.\ \cite{kaagstrom2008blocked} report that modest values between 5 and 12 are usually optimal.

After the multiplications from the left, the pencil is in a state shown in Figure~\ref{subfig:blockedstage1afterleftmul}. Now that the block column in $A$ has been reduced, we need to remove the fill-in in $B$ without perturbing the reduced columns of $A$. The simplest way to reduce $B$ back to upper triangular form would be to calculate the RQ factorizations of all the subblocks that were filled in and apply their orthogonal factors from the right. It was noted by K\r{a}gstr\"{o}m et al.\ \cite{kaagstrom2008blocked} that fully reducing $B$ back to upper triangular form is not necessary, only the first $n_b$ columns of the subblock need to be reduced.

Reducing a selection of columns using reflectors applied from the right is nontrivial. Usually, a reflector applied from the right reduces a selection of rows, not columns. To perform this task, opposite Householder reflectors can be employed. As first noted by D. Watkins \cite{watkins2000performance}, an opposite Householder reflector can reduce a column when applied from the right. To use this technique, we take the LQ factorization of the first $n_b$ rows of the orthogonal factor of the RQ factorization of the subblock. If we apply the orthogonal factor of this LQ factorization to $B$ from the right it will reduce the first $n_b$ columns of the subblock. For a detailed analysis, we refer to K\r{a}gstr\"{o}m et al.\ \cite{kaagstrom2008blocked}.

The way to generate the opposite reflectors may seem strange. We employ opposite reflectors to avoid the RQ decomposition of the subblock, but to generate the opposite reflectors we need to calculate the RQ decomposition anyway. The advantage becomes clear when we consider the cost of applying reflectors to the rest of the matrix. The RQ factorization uses $pn_b$ reflectors, whereas the LQ factorization only uses $n_b$ reflectors. The full reduction from the right is illustrated in Figure~\ref{subfig:blockedstage1rightmul}. Note that these multiplications do not affect the structure of $A$.

Figure~\ref{sub@subfig:blockedstage1afterrightmul} shows the final state of the pencil in this iteration. The subdiagonal blocks in $B$ are such that they will not interfere with subsequent iterations. During each iteration, the blocks in $B$ move down by $n_b$ positions eventually leading them off the edge of the pencil.

Finally, we consider the number of flops required for stage one. If we include the flops required to update $Q$ and $Z$, Algorithm~\ref{alg:blockedstage1} requires $\frac{28p+14}{3(p-1)}n^3 + O(n^2)$ flops. In our implementation, we have chosen $p=8$, which results in a cost of $11.33 \, n^3 + O(n^2)$ flops.

\subsection{Parallel stage one}

Because it is a blocked algorithm, Algorithm~\ref{alg:blockedstage1} already has some potential for parallelization. The application of a block reflector to a matrix uses several matrix-matrix multiplications, which can usually be parallelized well. This is essentially the parallelization strategy used by Dackland et al.\ \cite{dackland1998scalapack}. However, for typical choices of $r$ and $p$, the dimensions of these matrices are too small to effectively parallelize them. Here, we present a different parallelization of stage one.

We will be utilizing two levels of parallelization. First, we will identify several large-grained tasks. Second, we will split some of the large-grained tasks into fine-grained tasks.

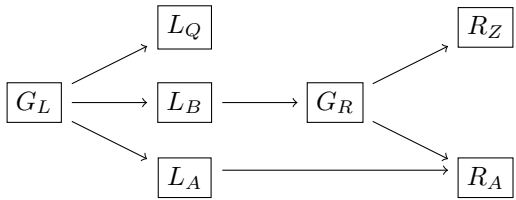
\begin{figure}[htb!]
	\centering
	\begin{tikzpicture}
		\node[draw] at (0,1) {$G_L$};

		\draw[->](0.5,0.75) -- (1.5,0.25);
		\draw[->](0.5,1.25) -- (1.5,1.75);
		\draw[->](0.5,1) -- (1.5,1);

		\node[draw] at (2,0) {$L_A$};
		\node[draw] at (2,1) {$L_B$};
		\node[draw] at (2,2) {$L_Q$};

		\draw[->](2.5,1) -- (3.5,1);

		\node[draw] at (4,1) {$G_R$};

		\draw[->](4.5,0.75) -- (5.5,0.25);
		\draw[->](4.5,1.25) -- (5.5,1.75);
		\draw[->](2.5,0.1) -- (5.5,0.1);

		\node[draw] at (6,0) {$R_A$};
		\node[draw] at (6,2) {$R_Z$};
	\end{tikzpicture}
	\caption{Dependency graph of parallel stage one.}
	\label{fig:dependecygraphstage1}
\end{figure}

Algorithm~\ref{alg:blockedstage1} operates in the following way: Calculate a block reflector and then apply this block reflector to the relevant matrices. For the updates from the left, an important realization is that the sequence of block reflectors depends only on the panel that is being reduced. We can introduce more parallelism by generating the sequence of reflectors as one task (the $G_L$ task) and then applying this sequence to $A$, $B$, and $Q$ in parallel (the $L_A$, $L_B$ and $L_Q$ tasks). For the updates from the right, the sequence of reflectors cannot be fully calculated without updating $B$. We can still split the work into the generation of the sequence of reflectors, which includes applying the sequence to $B$ (the $G_R$ task) and the application of this sequence to $A$ and $Z$ (the $R_A$ and $R_Z$ tasks). Figure~\ref{fig:dependecygraphstage1} illustrates the dependencies between the tasks.

\begin{figure}
	\centering
	\subcaptionbox{$L_A$}[.32\textwidth]{%
	\begin{tikzpicture}[scale=0.8]
		\foreach \i in {0,1,2,...,29}
			{
				\draw[color=lightgray,line width=0.1mm] (0.1*\i,0) -- (0.1*\i,3);
				\draw[color=lightgray,line width=0.1mm] (0,0.1*\i) -- (3,0.1*\i);
			}

		\draw (0,0) -- (0,3) -- (3,3) -- (3,0) -- (0,0);

		\fill[color=lightgray,opacity=0.8] (0.,1.0) -- (0.,0) -- (3,0) -- (3,1.0) -- (0.,1.0);
		\fill[color=lightgray,opacity=0.8] (0.,1.8) -- (0.,0.6) -- (3,0.6) -- (3,1.8) -- (0.,1.8);
		\fill[color=lightgray,opacity=0.8] (0.,2.6) -- (0.,1.4) -- (3,1.4) -- (3,2.6) -- (0.,2.6);

		\draw[fill=gray,fill opacity=0.8] (0,1.0) -- (0,0) -- (0.4,0) -- (0.4,0.6) -- (0.3,0.6) -- (0.3,0.7) -- (0.2,0.7) -- (0.2,0.8) -- (0.1,0.8) -- (0.1,0.9) -- (0.0,0.9);
		\draw[fill=gray,fill opacity=0.8] (0.0,1.0) -- (0.0,1.7) -- (0.1,1.7) -- (0.1,1.6) -- (0.2,1.6) -- (0.2,1.5) -- (0.3,1.5) -- (0.3,1.4) -- (0.4,1.4) -- (0.4,0.6) -- (0.3,0.6) -- (0.3,0.7) -- (0.2,0.7) -- (0.2,0.8) -- (0.1,0.8) -- (0.1,0.9) -- (0.0,0.9) -- (0.,1.0);
		\draw[fill=gray,fill opacity=0.8] (0.0,1.8) -- (0.0,2.5) -- (0.1,2.5) -- (0.1,2.4) -- (0.2,2.4) -- (0.2,2.3) -- (0.3,2.3) -- (0.3,2.2) -- (0.4,2.2) -- (0.4,1.4) -- (0.3,1.4) -- (0.3,1.5) -- (0.2,1.5) -- (0.2,1.6) -- (0.1,1.6) -- (0.1,1.7) -- (0.0,1.7) -- (0.,1.8);

		\draw [dashed] (0.4,0) -- (0.4, 3);
		\draw [dashed] (1.1,0) -- (1.1, 3);
		\draw [dashed] (1.8,0) -- (1.8, 3);
		\draw [dashed] (2.5,0) -- (2.5, 3);

		\node at (0.75,3.2) {$T_1$};
		\node at (1.45,3.2) {$T_2$};
		\node at (2.15,3.2) {$T_3$};
		\node at (2.75,3.2) {$T_4$};
	\end{tikzpicture}
	}
	\subcaptionbox{$R_A$}[.32\textwidth]{%
	\begin{tikzpicture}[scale=0.8]
		\foreach \i in {0,1,2,...,29}
			{
				\draw[color=lightgray,line width=0.1mm] (0.1*\i,0) -- (0.1*\i,3);
				\draw[color=lightgray,line width=0.1mm] (0,0.1*\i) -- (3,0.1*\i);
			}

		\draw (0,2.5) -- (0,3) -- (3,3) -- (3,0) -- (2.5,0);
		\foreach \i in {1,2,...,4}
			{
				\draw (0.1*\i,2.5-0.1*\i) -- (0.1*\i,2.5-0.1*\i+0.1) -- (0.1*\i-0.1,2.5-0.1*\i+0.1);
			}
		\draw (2.5,0) -- (0.4,0) -- (0.4,2.2);

		\fill[color=lightgray,opacity=0.8] (2.0,0.0) -- (3,0) -- (3,3) -- (2.0,3) -- (2.0,0);
		\fill[color=lightgray,opacity=0.8] (1.2,0) -- (2.4,0) -- (2.4,3) -- (1.2,3) -- (1.2,0);
		\fill[color=lightgray,opacity=0.8] (0.4,0) -- (1.6,0) -- (1.6,3) -- (0.4,3) -- (0.4,0);

		\draw [dashed] (0.4,0.8) -- (3, 0.8);
		\draw [dashed] (0.4,1.6) -- (3, 1.6);
		\draw [dashed] (0.4,2.4) -- (3, 2.4);

		\node at (3.2,2.7) {$T_1$};
		\node at (3.2,2.0) {$T_2$};
		\node at (3.2,1.2) {$T_3$};
		\node at (3.2,0.4) {$T_4$};

	\end{tikzpicture}
	}
	\subcaptionbox{$L_B$}[.32\textwidth]{%
	\begin{tikzpicture}[scale=0.8]
		\foreach \i in {0,1,2,...,29}
			{
				\draw[color=lightgray,line width=0.1mm] (0.1*\i,0) -- (0.1*\i,3);
				\draw[color=lightgray,line width=0.1mm] (0,0.1*\i) -- (3,0.1*\i);
			}

		\draw (0,2.9) -- (0,3) -- (3,3) -- (3,0) -- (2.9,0);
		\foreach \i in {1,2,...,29}
			{
				\draw (0.1*\i,2.9-0.1*\i) -- (0.1*\i,2.9-0.1*\i+0.1) -- (0.1*\i-0.1,2.9-0.1*\i+0.1);
			}

		\fill[color=lightgray,opacity=0.8] (2.0,1.0) -- (2.0,0) -- (3,0) -- (3,1.0) -- (2.0,1.0);
		\fill[color=lightgray,opacity=0.8] (1.2,1.8) -- (1.2,0.6) -- (3,0.6) -- (3,1.8) -- (1.2,1.8);
		\fill[color=lightgray,opacity=0.8] (0.4,2.6) -- (0.4,1.4) -- (3,1.4) -- (3,2.6) -- (0.4,2.6);

		\draw [dashed] (0.4,0) -- (0.4, 3);
		\draw [dashed] (1.1,0) -- (1.1, 3);
		\draw [dashed] (1.8,0) -- (1.8, 3);
		\draw [dashed] (2.5,0) -- (2.5, 3);

		\node at (0.75,3.2) {$T_1$};
		\node at (1.45,3.2) {$T_2$};
		\node at (2.15,3.2) {$T_3$};
		\node at (2.75,3.2) {$T_4$};

	\end{tikzpicture}
	}
	\caption{Illustration of the distribution of the $L_A$, $R_A$ and $L_B$ tasks of the parallel phase 1 algorithm to different subtasks $T_1,T_2,\dots$. The dark gray areas indicate the blocks that are reduced. The light gray areas indicate the parts of the matrix that are affected by the multiplications.}
	\label{fig:parallelstage1}
\end{figure}
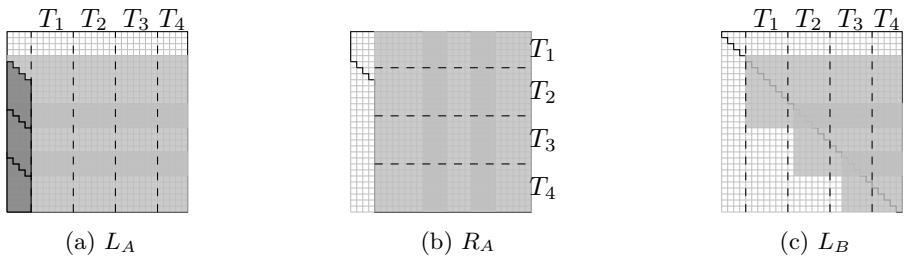

These large-grained tasks allow for some parallelism, but at most three tasks can be executed in parallel this way. To obtain an efficient algorithm, the tasks must be split into smaller tasks. The application tasks: $L_Q$, $L_A$ and $L_B$ apply a sequence of block reflectors to their respective matrices. To update a certain column of these matrices, the tasks only require that specific column, so $L_Q$, $L_A$ and $L_B$ can easily be parallelized by splitting the matrices into several column slices and updating the slices in parallel. The same trick can be applied to $R_Z$ and $R_A$ by splitting the matrices into row slices instead of column slices. The distribution is illustrated in Figure~\ref{fig:parallelstage1}. We emphasize that such a distribution of the matrices is probably not well suited to a distributed memory implementation, because the cores constantly switch between operating on rows and columns of the matrices. We also note that $L_B$ can have some load-balancing issues. Because of the upper-triangular structure of $B$, the different subtasks of $L_B$ consist of a different amount of flops. This could be solved by letting the tasks that affect fewer rows process more columns to equalize the number of flops. In our implementation, we chose to let the dynamic scheduler handle these load imbalances.

In a way, this parallelization is similar to the simple parallelization of the matrix-matrix multiplication mentioned at the start of this section. The difference is that instead of distributing the application of a single block reflector, we distribute the application of a sequence of block reflectors. This results in the same amount of parallelism, but there are fewer synchronization points. Finally, we consider the parallelization of the generation tasks: $G_L$ and $G_R$. $G_L$ is a small task and is not worth parallelizing. $G_R$ is the hardest task to parallelize. Calculating the next block reflector requires applying the previous block reflector to $B$, so the parallelization we used for $R_A$ is not applicable here. Only the simple parallelization of the matrix-matrix multiplications is possible.

\section{Reduction to Hessenberg-triangular form}
\label{sec: hessenberg-triangular}

To reduce a pencil in r-Hessenberg-triangular form to Hessenberg-triangular form, we adapt the reduction algorithm for r-Hessenberg matrices by Karlsson et al.\ \cite{karlsson2010efficient,karlsson2011parallel} to r-Hessenberg-triangular pencils. We start by explaining the unblocked algorithm and will explain how to delay and efficiently apply updates later.

\subsection{Unblocked stage two}\label{sec: unblockedstage2}

\begin{figure}
	\centering
	\subcaptionbox{Initial state of the pencil}[.49\textwidth]{
		\centering
		\begin{tikzpicture}[scale=2.5]

			\foreach \i in {0,1,2,...,10}
				{
					\draw[color=lightgray,line width=0.1mm] (0.1*\i,0) -- (0.1*\i,1);
					\draw[color=lightgray,line width=0.1mm] (0,0.1*\i) -- (1,0.1*\i);
				}

			\draw (0,0.6) -- (0,1) -- (1,1) -- (1,0) -- (0.6,0);
			\foreach \i in {1,2,...,6}
				{
					\draw (0.1*\i,0.6-0.1*\i) -- (0.1*\i,0.6-0.1*\i+0.1) -- (0.1*\i-0.1,0.6-0.1*\i+0.1);
				}

			\begin{scope}[shift={(1.1,0)}]
				\foreach \i in {0,1,2,...,10}
					{
						\draw[color=lightgray,line width=0.1mm] (0.1*\i,0) -- (0.1*\i,1);
						\draw[color=lightgray,line width=0.1mm] (0,0.1*\i) -- (1,0.1*\i);
					}
	
				\draw (0,0.9) -- (0,1) -- (1,1) -- (1,0) -- (0.9,0);
				\foreach \i in {1,2,...,9}
					{
						\draw (0.1*\i,0.9-0.1*\i) -- (0.1*\i,0.9-0.1*\i+0.1) -- (0.1*\i-0.1,0.9-0.1*\i+0.1);
					}

			\end{scope}
		\end{tikzpicture}
	}
	\subcaptionbox{Reducing a column in $A$\label{subfig:unblocked2-1}}[.49\textwidth]{
		\centering
		\begin{tikzpicture}[scale=2.5]

			\foreach \i in {0,1,2,...,10}
				{
					\draw[color=lightgray,line width=0.1mm] (0.1*\i,0) -- (0.1*\i,1);
					\draw[color=lightgray,line width=0.1mm] (0,0.1*\i) -- (1,0.1*\i);
				}

			\draw (0,0.6) -- (0,1) -- (1,1) -- (1,0) -- (0.6,0);
			\foreach \i in {1,2,...,6}
				{
					\draw (0.1*\i,0.6-0.1*\i) -- (0.1*\i,0.6-0.1*\i+0.1) -- (0.1*\i-0.1,0.6-0.1*\i+0.1);
				}

			\fill[color=lightgray,opacity=0.8] (0,0.6) -- (0,0.9) -- (1,0.9) -- (1,0.6) -- (0,0.6);
			\draw[fill=gray,fill opacity=0.8] (0,0.6) -- (0,0.8) -- (0.1,0.8) -- (0.1,0.6) -- (0,0.6);

			\begin{scope}[shift={(1.1,0)}]
				\foreach \i in {0,1,2,...,10}
					{
						\draw[color=lightgray,line width=0.1mm] (0.1*\i,0) -- (0.1*\i,1);
						\draw[color=lightgray,line width=0.1mm] (0,0.1*\i) -- (1,0.1*\i);
					}
	
				\draw (0,0.9) -- (0,1) -- (1,1) -- (1,0) -- (0.9,0);
				\foreach \i in {1,2,...,9}
					{
						\draw (0.1*\i,0.9-0.1*\i) -- (0.1*\i,0.9-0.1*\i+0.1) -- (0.1*\i-0.1,0.9-0.1*\i+0.1);
					}

				\fill[color=lightgray,opacity=0.8] (0.1,0.6) -- (0.1,0.9) -- (1,0.9) -- (1,0.6) -- (0.1,0.6);

			\end{scope}
		\end{tikzpicture}
	}
	\hfill
	\subcaptionbox{New state of the pencil}[.49\textwidth]{
		\centering
		\begin{tikzpicture}[scale=2.5]

			\foreach \i in {0,1,2,...,10}
				{
					\draw[color=lightgray,line width=0.1mm] (0.1*\i,0) -- (0.1*\i,1);
					\draw[color=lightgray,line width=0.1mm] (0,0.1*\i) -- (1,0.1*\i);
				}

			\draw (0,0.8) -- (0,1) -- (1,1) -- (1,0) -- (0.6,0);
			\foreach \i in {2,3,...,6}
				{
					\draw (0.1*\i,0.6-0.1*\i) -- (0.1*\i,0.6-0.1*\i+0.1) -- (0.1*\i-0.1,0.6-0.1*\i+0.1);
				}
			\draw (0,0.8) -- (0.1,0.8) -- (0.1,0.5);

			\begin{scope}[shift={(1.1,0)}]
				\foreach \i in {0,1,2,...,10}
					{
						\draw[color=lightgray,line width=0.1mm] (0.1*\i,0) -- (0.1*\i,1);
						\draw[color=lightgray,line width=0.1mm] (0,0.1*\i) -- (1,0.1*\i);
					}
	
				\draw (0,0.9) -- (0,1) -- (1,1) -- (1,0) -- (0.9,0);
				\foreach \i in {1,4,5,...,9}
					{
						\draw (0.1*\i,0.9-0.1*\i) -- (0.1*\i,0.9-0.1*\i+0.1) -- (0.1*\i-0.1,0.9-0.1*\i+0.1);
					}
				\draw (0.1,0.9) -- (0.1,0.6) -- (0.4,0.6);

			\end{scope}
		\end{tikzpicture}
	}
	\hfill
	\subcaptionbox{Reducing fill-in in $B$\label{subfig:unblocked2-2}}[.49\textwidth]{
		\centering
		\begin{tikzpicture}[scale=2.5]

			\foreach \i in {0,1,2,...,10}
				{
					\draw[color=lightgray,line width=0.1mm] (0.1*\i,0) -- (0.1*\i,1);
					\draw[color=lightgray,line width=0.1mm] (0,0.1*\i) -- (1,0.1*\i);
				}

			\draw (0,0.8) -- (0,1) -- (1,1) -- (1,0) -- (0.6,0);
			\foreach \i in {2,3,...,6}
				{
					\draw (0.1*\i,0.6-0.1*\i) -- (0.1*\i,0.6-0.1*\i+0.1) -- (0.1*\i-0.1,0.6-0.1*\i+0.1);
				}
			\draw (0,0.8) -- (0.1,0.8) -- (0.1,0.5);

			\fill[color=lightgray,opacity=0.8] (0.1,0.3) -- (0.1,1) -- (0.4,1) -- (0.4,0.3) -- (0.1,0.3);

			\begin{scope}[shift={(1.1,0)}]
				\foreach \i in {0,1,2,...,10}
					{
						\draw[color=lightgray,line width=0.1mm] (0.1*\i,0) -- (0.1*\i,1);
						\draw[color=lightgray,line width=0.1mm] (0,0.1*\i) -- (1,0.1*\i);
					}
	
				\draw (0,0.9) -- (0,1) -- (1,1) -- (1,0) -- (0.9,0);
				\foreach \i in {1,4,5,...,9}
					{
						\draw (0.1*\i,0.9-0.1*\i) -- (0.1*\i,0.9-0.1*\i+0.1) -- (0.1*\i-0.1,0.9-0.1*\i+0.1);
					}
				\draw (0.1,0.9) -- (0.1,0.6) -- (0.4,0.6);

				\fill[color=lightgray,opacity=0.8] (0.1,0.6) -- (0.1,1) -- (0.4,1) -- (0.4,0.6) -- (0.1,0.6);
				\draw[fill=gray,fill opacity=0.8] (0.1,0.6) -- (0.1,0.8) -- (0.2,0.8) -- (0.2,0.6) -- (0.1,0.6);

			\end{scope}
		\end{tikzpicture}
	}
	\hfill
	\subcaptionbox{Chase down fill-in\label{subfig:unblocked2-3}}[.49\textwidth]{
		\centering
		\begin{tikzpicture}[scale=2.5]

			\foreach \i in {0,1,2,...,10}
				{
					\draw[color=lightgray,line width=0.1mm] (0.1*\i,0) -- (0.1*\i,1);
					\draw[color=lightgray,line width=0.1mm] (0,0.1*\i) -- (1,0.1*\i);
				}

			\draw (0,0.8) -- (0,1) -- (1,1) -- (1,0) -- (0.6,0);
			\foreach \i in {4,5,6}
				{
					\draw (0.1*\i,0.6-0.1*\i) -- (0.1*\i,0.6-0.1*\i+0.1) -- (0.1*\i-0.1,0.6-0.1*\i+0.1);
				}
			\draw (0,0.8) -- (0.1,0.8) -- (0.1,0.3) -- (0.3,0.3);

			\fill[color=lightgray,opacity=0.8] (0.1,0.3) -- (1,0.3) -- (1,0.6) -- (0.1,0.6) -- (0.1,0.3);
			\fill[color=lightgray,opacity=0.8] (0.4,0) -- (0.4,1) -- (0.7,1) -- (0.7,0) -- (0.4,0);

			\draw[fill=gray,fill opacity=0.8] (0.1,0.3) -- (0.1,0.5) -- (0.2,0.5) -- (0.2,0.3) -- (0.1,0.3);

			\begin{scope}[shift={(1.1,0)}]
				\foreach \i in {0,1,2,...,10}
					{
						\draw[color=lightgray,line width=0.1mm] (0.1*\i,0) -- (0.1*\i,1);
						\draw[color=lightgray,line width=0.1mm] (0,0.1*\i) -- (1,0.1*\i);
					}
	
				\draw (0,0.9) -- (0,1) -- (1,1) -- (1,0) -- (0.9,0);
				\foreach \i in {4,5,...,9}
					{
						\draw (0.1*\i,0.9-0.1*\i) -- (0.1*\i,0.9-0.1*\i+0.1) -- (0.1*\i-0.1,0.9-0.1*\i+0.1);
					}
				\draw (0,0.9) -- (0.1,0.9) -- (0.1,0.8) -- (0.2,0.8) -- (0.2,0.6) -- (0.3,0.6);

				\fill[color=lightgray,opacity=0.8] (0.4,0.3) -- (1,0.3) -- (1,0.6) -- (0.4,0.6) -- (0.4,0.3);
				\fill[color=lightgray,opacity=0.8] (0.4,0.3) -- (0.4,1) -- (0.7,1) -- (0.7,0.3) -- (0.4,0.3);
			
				\draw[fill=gray,fill opacity=0.8] (0.4,0.3) -- (0.4,0.5) -- (0.5,0.5) -- (0.5,0.3) -- (0.4,0.3);

			\end{scope}
		\end{tikzpicture}
	}
	\hfill
	\subcaptionbox{Chase down fill-in\label{subfig:unblocked2-4}}[.49\textwidth]{
		\centering
		\begin{tikzpicture}[scale=2.5]

			\foreach \i in {0,1,2,...,10}
				{
					\draw[color=lightgray,line width=0.1mm] (0.1*\i,0) -- (0.1*\i,1);
					\draw[color=lightgray,line width=0.1mm] (0,0.1*\i) -- (1,0.1*\i);
				}

			\draw (0,0.8) -- (0,1) -- (1,1) -- (1,0) -- (0.6,0);
			\draw (0,0.8) -- (0.1,0.8) -- (0.1,0.5) -- (0.2,0.5) -- (0.2,0.3) -- (0.4,0.3) -- (0.4,0) -- (0.6,0);

			\fill[color=lightgray,opacity=0.8] (0.4,0) -- (1,0) -- (1,0.3) -- (0.4,0.3) -- (0.4,0);
			\fill[color=lightgray,opacity=0.8] (0.7,0) -- (0.7,1) -- (1,1) -- (1,0) -- (0.7,0);

			\draw[fill=gray,fill opacity=0.8] (0.4,0) -- (0.4,0.2) -- (0.5,0.2) -- (0.5,0) -- (0.4,0);

			\begin{scope}[shift={(1.1,0)}]
				\foreach \i in {0,1,2,...,10}
					{
						\draw[color=lightgray,line width=0.1mm] (0.1*\i,0) -- (0.1*\i,1);
						\draw[color=lightgray,line width=0.1mm] (0,0.1*\i) -- (1,0.1*\i);
					}
	
				\draw (0,0.9) -- (0,1) -- (1,1) -- (1,0) -- (0.9,0);
				\foreach \i in {4,7,8,9}
					{
						\draw (0.1*\i,0.9-0.1*\i) -- (0.1*\i,0.9-0.1*\i+0.1) -- (0.1*\i-0.1,0.9-0.1*\i+0.1);
					}
				\draw (0,0.9) -- (0.1,0.9) -- (0.1,0.8) -- (0.2,0.8) -- (0.2,0.6) -- (0.3,0.6);
				\draw (0.4,0.5) -- (0.5,0.5) -- (0.5,0.3) -- (0.6,0.3);

				\fill[color=lightgray,opacity=0.8] (0.7,0) -- (1,0) -- (1,0.3) -- (0.7,0.3) -- (0.7,0);
				\fill[color=lightgray,opacity=0.8] (0.7,0) -- (0.7,1) -- (1,1) -- (1,0) -- (0.7,0);

				\draw[fill=gray,fill opacity=0.8] (0.7,0) -- (0.7,0.2) -- (0.8,0.2) -- (0.8,0) -- (0.7,0);
			\end{scope}
		\end{tikzpicture}
	}
	\caption{Illustration of Algorithm~\ref{alg:unblockedstage2}.}
	\label{fig:illustrationunblockedstage2}
\end{figure}
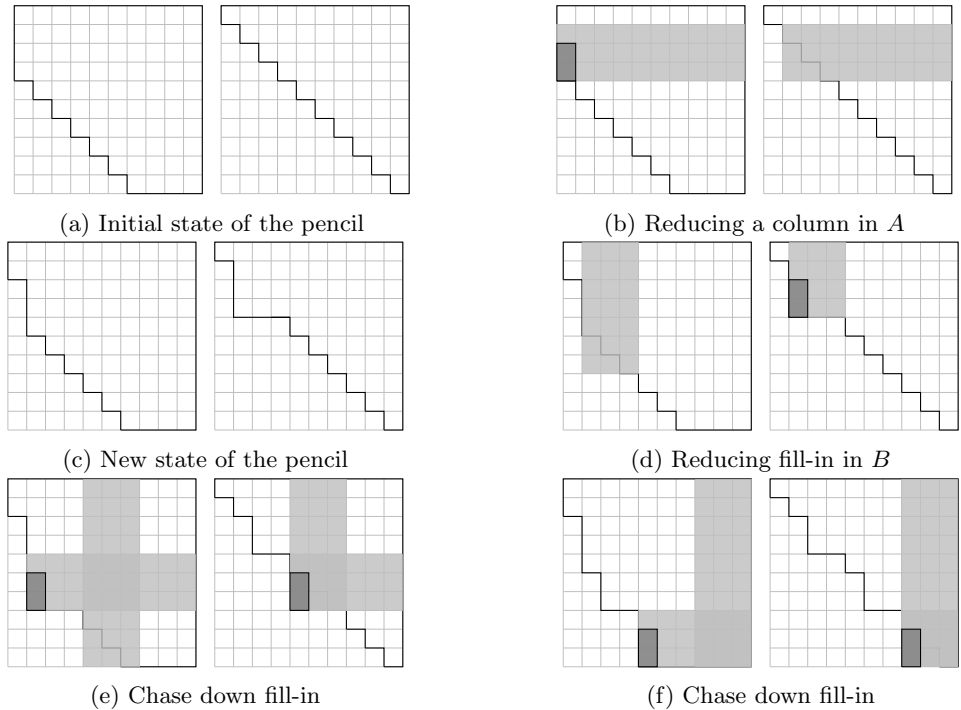

Just like in stage one, we will reduce entries in $A$ with reflectors applied from the left, while eliminating the fill-in in $B$ with reflectors applied from the right. The algorithm consists of multiple sweeps. During one sweep, we will reduce one column in $A$ and eliminate the fill-in. At the start of sweep $j$, the first $j-1$ columns are already in Hessenberg-triangular form. The first step in a sweep is to apply a reflector $\hat{Q}_0^j$ from the left so that $\hat{Q}_0^jA(j+1:j+r,j)$ is reduced. This will create fill-in in $B$ so that there is a block in $B(j+1:j+r,j+1:j+r)$. This first step is illustrated in Figure~\ref{subfig:unblocked2-1}. Next, we apply an opposite reflector $\hat{Z}_0^j$ so that the first column of $B(j+1:j+r,j+1:j+r)\hat{Z}_0^j$ is reduced. This second step is illustrated in Figure~\ref{subfig:unblocked2-2}. Notice that $\hat{Z}_0^j$ creates some fill-in in $A$. We can eliminate this fill-in using $\hat{Q}_1^j$ and eliminate the resulting fill-in in $B$ using $\hat{Z}_1^j$. As Figures~\ref{subfig:unblocked2-3} and \ref{subfig:unblocked2-4} illustrate, this leads to a bulge-chasing algorithm where eliminating the fill-in (sometimes called a bulge) leads to more fill-in further down the matrix. This process is repeated until the bulge is chased off the end of the matrix. The full process is shown in Algorithm~\ref{alg:unblockedstage2}.

Note that just like in stage one $B$ is block-upper-triangular (and not upper triangular) after the first sweep. Similarly, $A$ is not in r-Hessenberg-triangular form. These blocks do not interfere with subsequent sweeps and move down as more columns are reduced. Traditionally, stage two for Hessenberg-triangular matrices has been implemented using Givens rotations, a description of this algorithm can be found in the work of K{\aa}gstr{\"o}m et al.\ \cite{kaagstrom2008blocked}. Givens rotations avoid the RQ factorizations required for calculating the opposite reflectors, but they (asymptotically) require slightly more flops.

If we include the flops required to update $Q$ and $Z$, Algorithm~\ref{alg:unblockedstage2} requires $10n^3 + O(n^2)$ flops. Note that the $r^2n^2$ cost of the RQ factorizations can be significant if $r$ is large even if it is asymptotically negligible. In our implementation, we usually choose $r = 16$. The original reduction algorithm of Moler and Stewart requires only $14n^3 + O(n^2)$ flops, while the two-stage reduction requires $21.33 n^3 + O(n^2)$ flops. That is an increase of more than $40\%$. In the right circumstances, parallelization will compensate for the increased computational cost.

\subsection{Blocked stage two}\label{sec: blockedstage2}

Our blocked stage two algorithm has a simple premise: update as few entries as possible while generating the reflectors and then use the additional freedom to reorder the operations to achieve higher efficiency when fully applying the updates. We will start by explaining how to reorder and apply the updates and end by showing how to generate the reflectors. These are respectively the application and generate phases of the blocked algorithm.

\paragraph{Applying the reflectors}

Assume for the moment that we can somehow generate the reflectors for $q$ consecutive columns without updating the entire matrix. Even in this case, efficiently applying the reflectors is nontrivial. The unblocked variant applies the reflectors in the following order
\begin{equation*}
	\hat{Q}_0^j,\hat{Z}_0^j,\hat{Q}_{1}^j,\hat{Z}_1^j,\dots,\hat{Q}_0^{j+1},\hat{Z}_0^{j+1},\hat{Q}_{1}^{j+1},\hat{Z}_{1}^{j+1},\dots
\end{equation*}
An important realization is that if we already have all the reflectors we can apply all the $\hat{Z}$ first and apply the $\hat{Q}$ later because of associativity. This leads to the following order
\begin{equation*}
	\hat{Z}_0^j,\hat{Z}_{1}^j,\dots,\hat{Z}_0^{j+1},\hat{Z}_{1}^{j+1},\dots,\hat{Q}_0^j,\hat{Q}_{1}^j,\dots,\hat{Q}_0^{j+1},\hat{Q}_{1}^{j+1},\dots
\end{equation*}
One might hope that because the updates from the left and right are now separated, applying them will be more efficient. Unfortunately, there is not much cache reuse. The entries affected by $\hat{Q}_k^j$ do not overlap with the entries affected by $\hat{Q}_{k+1}^{j}$, so there is almost no cache reuse if the reflectors are applied in this order. We have some additional freedom in choosing the order. Reflectors that do not overlap commute. This ultimately means that as long as a reflector $\hat{Q}_k^j$ is preceded by $\hat{Q}_k^{j-1}$ and $\hat{Q}_{k+1}^{j-1}$ (and similarly for $\hat{Z}$) we can choose any order we want. A reflector $\hat{Q}_k^j$ affects $r$ columns (or rows in the case of $\hat{Z}_k^j$), $r-1$ of which are shared with $\hat{Q}_k^{j+1}$, so there is much more potential for cache reuse. Grouping the reflectors by $k$ instead of $j$ is likely more efficient. Consider the following sequence
\begin{equation*}
	\hat{Z}_{l-1}^j,\hat{Z}_{l-1}^{j+1},\dots,\hat{Z}_{l-2}^{j},\hat{Z}_{l-2}^{j+1},\dots,\hat{Q}_{l-1}^j,\hat{Q}_{l-1}^{j+1},\dots,\hat{Q}_{l-2}^{j},\hat{Q}_{l-2}^{j+1},\dots,
\end{equation*}
where $l$ is the maximal value of $k$. It is easy to see that the reflectors are now grouped by $k$ and that the previously mentioned constraint is not violated. This technique was initially discovered by C. Bishof, X. Sun, and B. Lang \cite{bischof1994parallel} for tridiagonal reductions. Processing the reflectors in this order also allows us to use WY representations. Each group of $q$ reflectors belonging to the same $k$ can be accumulated into a block reflector so that we can use matrix-matrix multiplications.

\paragraph{Generating the reflectors}

Now that we know how to efficiently apply the reflectors, we need a way to calculate the reflectors for several sweeps while updating as few entries as possible. This amounts to the following task. Generate the reflectors for the columns $j = j_1:j_1+q-1$, while only applying the reflectors $\hat{Q}_k^j$ to a range of columns $c_{1A}(k,j):c_{2A}(k,j)$ of $A$ and columns $c_{1B}(k,j):c_{2B}(k,j)$ of $B$ and applying the reflectors $\hat{Z}_k^j$ to a range of rows $r_{1A}(k,j):r_{2A}(k,j)$ of $A$ and rows $r_{1B}(k,j):r_{2B}(k,j)$ of $B$.

Let us start by deriving expressions for $A$. In order to generate the reflector $\hat{Q}_k^{j+1}$,
\begin{equation}
	r_{1A}(k,j) \le j_1 + k r + 1 - r
\end{equation}
and in order to generate the reflector $\hat{Q}_{k+1}^{j}$, 
\begin{equation}
	r_{2A}(k,j) \ge \min(j+(k+2)r,n).
\end{equation}
We must also respect the constraints that were previously mentioned, $Z_k^j$ must be preceded by $Z_k^{j-1}$ and $Z_{k+1}^{j-1}$, this means that
\begin{equation}
	r_{1A}(k,j) \ge r_{1A}(k,j-1)
\end{equation}
and
\begin{equation*}\label{eq:r1}
	r_{1A}(k,j) \ge r_{1A}(k-1,j-1)
\end{equation*}
Setting $r_{1A}(k,j_1+q-1) = j_1 + k r + 1 - r$ and combining that with equation~\eqref{eq:r1} we get that $r_{1A}(k,j_1+q-2) = j_1 + k r + 1 - 2r$. We can continue this logic to find that $r_{1A}(k,j_1+q-3) = j_1 + k r + 1 - 3r$ and we ultimately get
\begin{equation}
	\begin{split}
		r_{1A}(k,j) &= j_1 + 1 + \max(0,kr - r - (j_1+q-1 - j )r)\\
		r_{2A}(k,j) &= \min(j+(k+2)r,n).
	\end{split}
\end{equation}

For $Q_k^j$, the range is simpler, we can simply set
\begin{equation*}
	c_{1A}(k,j) = j + \max(0,(k-1)r + 1)
\end{equation*}
and
\begin{equation*}
	c_{2A}(k,j) = j_1 + q - 1 + \max(0,(k-1)r + 1)
\end{equation*}
These are the minimal ranges required for generating the reflectors and it does not need to be expanded to account for overlapping updates. These two ranges were first presented by L. Karlsson et al.\ \cite{karlsson2010efficient} for Hessenberg matrices.

A similar derivation for $B$ results in:
\begin{equation}
	\begin{split}
		r_{1B}(k,j) &= j_1 + 1 + \max(0,kr - r - (j_1+q-1 - j )r)\\
		r_{2B}(k,j) &= \min(j+(k+1)r,n)
	\end{split}
\end{equation}
and
\begin{equation}
	\begin{split}
		c_{1B}(k,j) &= j+kr+1\\
		c_{2B}(k,j) &= \min(j_1 + q - 1 +(k+1)r,n).
	\end{split}
\end{equation}

Algorithm~\ref{alg:stage2generate} generates the reflectors for a sequence of columns while only updating a small part of $A$ and $B$. We invite the reader to study the differences between Algorithm~\ref{alg:unblockedstage2} and Algorithm~\ref{alg:stage2generate}. Algorithm~\ref{alg:stage2application} applies the rest of the updates using reordered reflectors and block reflectors. Figures~\ref{fig:stage2generateleft} and \ref{fig:stage2generateright} show the parts of the pencil that are updated from the left and right while generating the reflectors.

\input{Figures/stage2generate.tex}

\subsection{Parallel stage two}

The parallelization strategy of stage two follows the same general principle as stage one. We split the generation of the reflectors and the updates of the different matrices into different tasks. What is special here is that we also split the application tasks into lookahead tasks and smaller application tasks. During these lookahead tasks, we update $A$ and $B$ just enough so that the $O(r q)$ band required for the generation of the reflectors is fully updated. This allows us to overlap the updating of the matrices with the generation of the reflectors. This is important because the generation of the reflectors is neither negligible nor easily parallelizable. Figure~\ref{fig:dependecygraphstage2} illustrates the dependencies between the tasks of different iterations.

Next, we parallelize the large tasks in the same way as in stage one. For both the lookahead and update tasks, we split the matrices into either column or row slices and apply the updates in parallel. Figure~\ref{fig: parallelphase2} shows this distribution.

\begin{figure}[htb!]
	\centering
	\begin{tikzpicture}[scale=0.8, every node/.style={scale=0.85}]
		\node[draw] at (-0.5,1) {$\text{Generate}^{(1)}$};

		\draw[->](0.6,0.75) -- (1.35,0.25);
		\draw[->](0.6,1) -- (1.35,1);
		\draw[->](0.6,1.125) -- (3.25,3.75);
		\draw[->](0.6,1.35) -- (3.25,4.75);

		\node[draw] at (3,0) {Lookahead $A^{(1)}$};
		\node[draw] at (3,1) {Lookahead $B^{(1)}$};
		\node[draw] at (4.5,4) {Update $Q^{(1)}$};
		\node[draw] at (4.5,5) {Update $Z^{(1)}$};

		\draw[->](4.5,1.25) -- (4.8,2.75);
		\draw[->](4.5,0.125) -- (4.8,1.75);

		\node[draw] at (6,2) {Update $A^{(1)}$};
		\node[draw] at (6,3) {Update $B^{(1)}$};

		\draw[->](6.5,4) -- (9,4);
		\draw[->](6.5,5) -- (9,5);
		\draw[->](7.2,3) -- (8,1);
		\draw[->](7.2,2) -- (8,0.25);

		\draw[->](4.5,0) -- (4.8,1);
		\draw[->](4.5,1) -- (4.8,1);

		\begin{scope}[shift={(6.5,0)}]
			\node[draw] at (-0.5,1) {$\text{Generate}^{(2)}$};
	
			\draw[->](0.6,0.75) -- (1.35,0.25);
			\draw[->](0.6,1) -- (1.35,1);
			\draw[->](0.6,1.125) -- (3.25,3.75);
			\draw[->](0.6,1.35) -- (3.25,4.75);
	
			\node[draw] at (3,0) {Lookahead $A^{(2)}$};
			\node[draw] at (3,1) {Lookahead $B^{(2)}$};
			\node[draw] at (4.5,4) {Update $Q^{(2)}$};
			\node[draw] at (4.5,5) {Update $Z^{(2)}$};
	
			\draw[->](4.5,1.25) -- (4.8,2.75);
			\draw[->](4.5,0.125) -- (4.8,1.75);

			\node[draw] at (6,2) {Update $A^{(2)}$};
			\node[draw] at (6,3) {Update $B^{(2)}$};
		\end{scope}




	\end{tikzpicture}
	\caption{Dependency graph of parallel stage two. The superscript indicates which iteration the task belongs to.}
	\label{fig:dependecygraphstage2}
\end{figure}
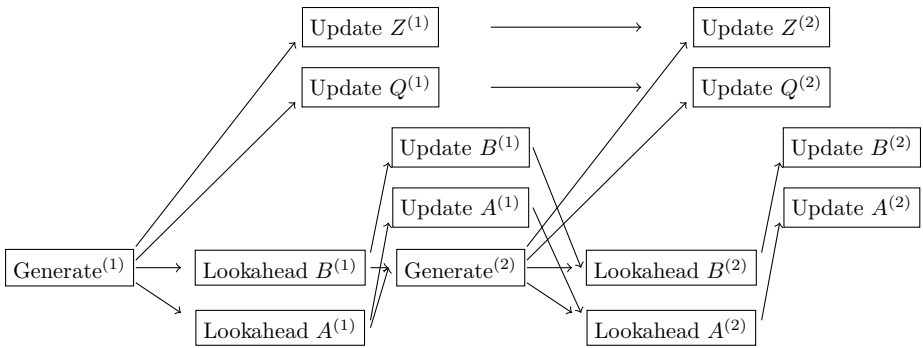

\begin{figure}[htb]
	\centering
	\subcaptionbox{full updates from the right and generation of reflectors}[.49\textwidth]{%
		\begin{tikzpicture}[scale=1]

			\draw (0,2.7) -- (0,3) -- (3,3) -- (3,0) -- (2.7,0) -- (0,2.7);
         \draw[fill=lightgray] (0.6,3) -- (3,3) -- (3,0.6) -- (0.6,3);
         \draw[fill=darkgray] (0,2.7) -- (0,3) -- (0.6,3) -- (3,0.6) -- (3,0) -- (2.7,0) -- (0,2.7);

			\foreach \i in {1,2,3,4,5,6,7}
				{
					\draw[dashed] (0.6+0.3*\i,3-0.3*\i) -- (3,3-0.3*\i);
				}
						
         \begin{scope}[shift={(3.3,0)}]
	
				\draw (0,3) -- (3,3) -- (3,0) -- (0,3);

				\draw[fill=lightgray] (0.6,3) -- (3,3) -- (3,0.6) -- (0.6,3);
         	\draw[fill=darkgray] (0,3) -- (0.6,3) -- (3,0.6) -- (3,0) -- (0,3);

				\foreach \i in {1,2,3,4,5,6,7}
					{
						\draw[dashed] (0.6+0.3*\i,3-0.3*\i) -- (3,3-0.3*\i);
					}

			\end{scope}
		\end{tikzpicture}
	}
	\subcaptionbox{full updates from the left and generation of reflectors}[.49\textwidth]{%
		\begin{tikzpicture}[scale=1]

			\draw (0,2.7) -- (0,3) -- (3,3) -- (3,0) -- (2.7,0) -- (0,2.7);
         \draw[fill=lightgray] (0.6,3) -- (3,3) -- (3,0.6) -- (0.6,3);
         \draw[fill=darkgray] (0,2.7) -- (0,3) -- (0.6,3) -- (3,0.6) -- (3,0) -- (2.7,0) -- (0,2.7);

			\foreach \i in {1,2,3,4,5,6,7}
				{
					\draw[dashed] (0.6+0.3*\i,3) -- (0.6+0.3*\i,3-0.3*\i);
				}
						
         \begin{scope}[shift={(3.3,0)}]
	
				\draw (0,3) -- (3,3) -- (3,0) -- (0,3);

				\draw[fill=lightgray] (0.6,3) -- (3,3) -- (3,0.6) -- (0.6,3);
         	\draw[fill=darkgray] (0,3) -- (0.6,3) -- (3,0.6) -- (3,0) -- (0,3);

				\foreach \i in {1,2,3,4,5,6,7}
					{
						\draw[dashed] (0.6+0.3*\i,3) -- (0.6+0.3*\i,3-0.3*\i);
					}

			\end{scope}
		\end{tikzpicture}
	}
	\subcaptionbox{lookahead updates from the right}[.49\textwidth]{%
		\begin{tikzpicture}[scale=1]

			\draw (0,2.7) -- (0,3) -- (3,3) -- (3,0) -- (2.7,0) -- (0,2.7);
         \draw[fill=gray] (0.3,3) -- (1.2,3) -- (3,1.2) -- (3,0.3) -- (0.3,3);

			\foreach \i in {2,4,6}
				{
					\draw[dashed] (0.3+0.3*\i,3-0.3*\i) -- (1.2+0.3*\i,3-0.3*\i);
				}
						
         \begin{scope}[shift={(3.3,0)}]
	
				\draw (0,3) -- (3,3) -- (3,0) -- (0,3);

				\draw[fill=gray] (0.3,3) -- (1.2,3) -- (3,1.2) -- (3,0.3) -- (0.3,3);

				\foreach \i in {2,4,6}
				{
					\draw[dashed] (0.3+0.3*\i,3-0.3*\i) -- (1.2+0.3*\i,3-0.3*\i);
				}

			\end{scope}
		\end{tikzpicture}
	}
	\subcaptionbox{lookahead updates from the left}[.49\textwidth]{%
		\begin{tikzpicture}[scale=1]

			\draw (0,2.7) -- (0,3) -- (3,3) -- (3,0) -- (2.7,0) -- (0,2.7);
         \draw[fill=gray] (0.3,3) -- (1.2,3) -- (3,1.2) -- (3,0.3) -- (0.3,3);

			\foreach \i in {3,5,7}
				{
					\draw[dashed] (0.3+0.3*\i,3.9-0.3*\i) -- (0.3+0.3*\i,3-0.3*\i);
				}
						
         \begin{scope}[shift={(3.3,0)}]
	
				\draw (0,3) -- (3,3) -- (3,0) -- (0,3);

				\draw[fill=gray] (0.3,3) -- (1.2,3) -- (3,1.2) -- (3,0.3) -- (0.3,3);

				\foreach \i in {3,5,7}
				{
					\draw[dashed] (0.3+0.3*\i,3.9-0.3*\i) -- (0.3+0.3*\i,3-0.3*\i);
				}

			\end{scope}
		\end{tikzpicture}
	}
   \caption{Distribution of the update, lookahead and generate tasks of phase 2 into multiple subtasks. In dark gray, the generate task, this task cannot be parallelized effectively and is not split into subtasks. The full update tasks are shown in light gray. They encompass most of the flops of the algorithm and are split into the most subtasks. The lookahead updates are shown in gray. These tasks encompass fewer flops so they are split into larger blocks.}
	\label{fig: parallelphase2}
\end{figure}

\section{Experiments}\label{sec: experiments}

We compare the following implementations
\begin{itemize}
	\item ParaHT: The algorithm presented in this paper
	\item IterHT: Iterative reduction algorithm by T. Steel and R. Vandebril \cite{steel2023}
	\item DGGHD3: The blocked Hessenberg-triangular reduction by K\r{a}gstr\"{o}m  et al.\ \cite{kaagstrom2008blocked} as implemented in LAPACK 3.9 \cite{laug}.
	\item HouseHT: Householder reflector based reduction by Bujanovic, L. Karlsson and D. Kressner \cite{bujanovic2018householder}.
\end{itemize}

The tests are performed on a machine with two Intel Xeon E-2697 v3 CPUs (14 cores each) that share 128GB of memory. A single node of this CPU has a theoretical peak double precision flop rate of 20.8 Gflops.

ParaHT, IterHT, and DGGHD3 are written in Fortran and compiled with GNU Fortran version 7.5.0 with optimization flags \verb|-O3 -march=native|. HouseHT is written in C++ and compiled with GNU C Compiler version 7.5.0 with optimization flags \verb|-O3 -march=native|. The code is linked with MKL version 2019.0.1.

For HouseHT, the parameter $n_b$ is set to 64. For ParaHT, the parameters $r$, $p$, and $q$ are set to 16, 8, and 8 respectively. These parameters were tuned by running the algorithm on randomly generated pencils of different sizes and choosing the set of parameters that performed the best on average.

The experiments will focus on the runtime of the algorithms. We will not go into detail about the accuracy of the methods. Each of the algorithms that we tested always produces results that have relative backward errors on the order of the machine precision.

\paragraph*{Tests on random pencils}

To evaluate the general performance of our algorithm, we run it for randomly generated matrices of varying sizes and compare it with other implementations. After generating the random pencil, we take a QR factorization of $B$ so that it is upper triangular. Note that a random matrix is usually well conditioned, which is important for two of the algorithms we compare to. IterHT and to a lesser degree HouseHT are iterative algorithms that require more iterations if the pencil has many infinite eigenvalues. Pencils of which the matrix $B$ has a small condition number avoid these extra iterations. 

Figure~\ref{subfig:parallelspeedup1} shows the runtime of several algorithms relative to the sequential runtime of LAPACK for a varying number of threads. Because of the extra flops, our algorithm is significantly slower than the other algorithms on one core. As the number of threads increases, the effective parallelization makes up for the increased cost. We also note that HouseHT and IterHT achieve their highest parallel speedup when using 14 threads. In other tests that do not explicitly vary the number of threads, we limit HouseHT and IterHT to 14 threads to get a fair comparison.

Figure~\ref{subfig:varyingsizespeedup} shows the speedup our algorithm achieves over LAPACK, HouseHT and IterHT. We achieve a speedup of 2 over HouseHT. Against LAPACK, our algorithm is slightly slower for small matrices but achieves a speedup of 4 for large matrices. IterHT is faster than our algorithm in most cases. As mentioned before, the runtime of IterHT strongly depends on the number of infinite eigenvalues present in the pencil and the condition of $B$. In most of the tests, because $B$ is well-conditioned, it only needs a single iteration. For $n=8000$ and $n=16000$, it needs two iterations and in those cases, our algorithm is slightly faster than IterHT.

Figure~\ref{fig:phase1phase2comparison} shows the parallel performance of ParaHT in more detail. First, note that most of the runtime of the algorithm is spent inside phase 2 despite phase 1 requiring slightly more flops. This indicates that phase 2 is less efficient. Second, the parallel speedup of phase 1 and phase 2 (and consequently, the full algorithm) is very similar. The issues that make phase 2 less efficient on a single core persist when executing the phases in parallel. Finally, we can also see that the matrices need to be quite large to achieve good speedups. For $n=1000$, our algorithm only achieves a parallel speedup of around 2, but for $n=8000$, our algorithm achieves a parallel speedup of 10.

\begin{figure}[htb]
	\centering

	\subcaptionbox{Parallel speedup\label{subfig:parallelspeedup1}}[.49\textwidth]{%
		\begin{tikzpicture}
			\begin{loglogaxis}[
					xlabel=Number of threads,
					ylabel=Parallel speedup,
					legend style={at={(0.03,0.97)},anchor=north west,nodes={scale=0.9, transform shape}},
					ytick={0.5,1,2,4,6,8},
					yticklabels={0.5,1,2,4,6,8},
					ymax=10,
					xtick={1,2,4,8,14,28},
					xticklabels={1,2,4,8,14,28},
					grid=major,
					width=0.5\textwidth,
					height=0.5\textwidth,
				]

				\pgfplotstableread[row sep=\\,col sep=&]{
					nthread & tlapack      & thouseht       & titerht       & tmdk\\
					1       & 649.361360   & 600.232372     & 621.098111    & 1183.129275 \\
					2       & 597.890039   & 467.141635     & 371.982559    & 744.187196  \\
					4       & 517.435487   & 354.648676     & 217.396199    & 403.227167  \\
					8       & 491.315822   & 291.794363     & 168.089519    & 186.901355  \\
					14      & 469.670329   & 284.524092     & 151.129816    & 126.416005   \\
					28      & 489.979042   & 611.756657     & 167.637667    & 112.244910   \\
				}\mydata

				\addplot[mark=*,black] table [x=nthread, y expr=649.361360/\thisrow{tlapack}] {\mydata};
				\addplot[mark=square,blue] table [x=nthread, y expr=649.361360/\thisrow{thouseht}] {\mydata};
				\addplot[dashed,mark=*,mark options={solid},red] table [x=nthread, y expr=649.361360/\thisrow{titerht}] {\mydata};
				\addplot[dashed,mark=square,mark options={solid},purple] table [x=nthread, y expr=649.361360/\thisrow{tmdk}] {\mydata};

				\legend{LAPACK\\HouseHT\\IterHT\\ParaHT\\}

			\end{loglogaxis}
		\end{tikzpicture}
	}
	\subcaptionbox{Speedup for varying n\label{subfig:varyingsizespeedup}}[.49\textwidth]{%
		\begin{tikzpicture}
			\begin{loglogaxis}[
					xlabel=$n$,
					ylabel=Speedup,
					legend style={at={(0.03,0.97)},anchor=north west,nodes={scale=0.9, transform shape}},
					ytick={0.5,1,2,4,8},
					yticklabels={0.5,1,2,4,8},
					xtick={500,1000,2000,4000,8000,16000},
					xticklabels={500,1000,2000,4000,8000,16000},
					ymax=12,
					grid=major,
					width=0.5\textwidth,
					height=0.5\textwidth,
				]

				\pgfplotstableread[row sep=\\,col sep=&]{
					n       & tlapack       & thouseht      & titerht      & tmdk\\
					500     & 0.145698      & 0.690084      & 0.133526     & 0.200381  \\
					707     & 0.369715      & 1.058591      & 0.251812     & 0.384931  \\
					1000    & 0.956220      & 2.080904      & 0.437296     & 0.738822  \\
					1414    & 2.528814      & 3.501035      & 0.761476     & 1.530092  \\
					2000    & 7.510552      & 7.228811      & 1.453135     & 3.538776   \\
					2828    & 21.422689     & 16.895852     & 2.882595     & 7.807473    \\
					4000    & 57.865404     & 41.768982     & 8.310056     & 19.043631    \\
					5657    & 168.395308    & 107.351477    & 26.809800    & 47.427194    \\
					8000    & 479.150683    & 293.304105    & 160.185686   & 126.563742    \\
					11314   & 1415.677491   & 782.748799    & 257.590574   & 318.964319    \\
					16000   & 4098.622276   & 2294.581108   & 1482.026003  & 962.339529    \\
				}\mydata

				\addplot[mark=*,black] table [x=n, y expr=\thisrow{tlapack}/\thisrow{tmdk}] {\mydata};
				\addplot[mark=square,blue] table [x=n, y expr=\thisrow{thouseht}/\thisrow{tmdk}] {\mydata};
				\addplot[dashed,mark=*,mark options={solid},red] table [x=n, y expr=\thisrow{titerht}/\thisrow{tmdk}] {\mydata};

				\legend{LAPACK\\HouseHT\\IterHT\\}

			\end{loglogaxis}
		\end{tikzpicture}
	}
	\caption{On the left, the parallel speedup (fraction of single-threaded runtime of LAPACK and multi-threaded runtime of the relevant algorithm) for a randomly generated pencil of size 8000. On the right, the speedup ParaHT achieves over other algorithms for varying pencil sizes.}
	\label{fig:algorithmcomparison}
\end{figure}
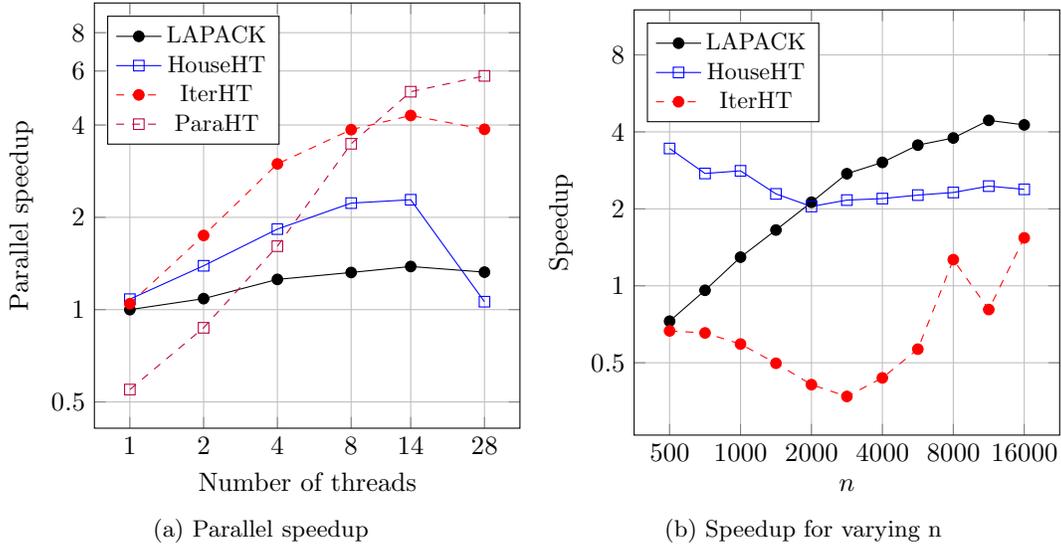

\begin{figure}[htb]
	\centering
	\subcaptionbox{Runtime of phase 1 and 2 relative to the total runtime
	\label{subfig:relativeruntimephase1}}[.49\textwidth]{%
		\begin{tikzpicture}
			\begin{semilogxaxis}[
					xlabel=Number of threads,
					ylabel=Relative runtime,
					legend style={at={(0.03,0.5)},anchor=north west,nodes={scale=0.9, transform shape}},
					xtick={1,2,4,8,14,28},
					xticklabels={1,2,4,8,14,28},
					grid=major,
					width=0.5\textwidth,
				]

				\pgfplotstableread[row sep=\\,col sep=&]{
					nthread & t1000    & t1000phase1 & t1000phase2 & t2000     & t2000phase1 & t2000phase2 & t4000      & t4000phase1 & t4000phase2 & t8000       & t8000phase1 & t8000phase2  \\
					1       & 1.678905 & 0.621871    & 1.057035    & 14.948054 & 5.650263    & 9.297791    & 132.776269 & 46.639628   & 86.136641   & 1183.129275 & 386.106099  & 797.023176  \\
					2       & 1.017018 & 0.415308    & 0.601710    & 7.863060  & 2.834056    & 5.029004    & 73.398768  & 26.542786   & 46.855981   & 744.187196  & 249.537004  & 494.650192  \\
					4       & 0.845592 & 0.357868    & 0.487725    & 4.816636  & 2.140299    & 2.676337    & 40.209688  & 15.154549   & 25.055139   & 403.227167  & 134.187146  & 269.040021  \\
					8       & 0.795709 & 0.322723    & 0.472986    & 3.723799  & 1.593960    & 2.129839    & 22.943964  & 10.004285   & 12.939679   & 186.901355  & 69.896575   & 117.004780  \\
					14      & 0.750277 & 0.298037    & 0.452240    & 3.571014  & 1.447620    & 2.123394    & 18.832407  & 7.951350    & 10.881056   & 126.416005  & 51.091414   & 75.324591    \\
					28      & 0.751109 & 0.300108    & 0.451001    & 3.524856  & 1.382442    & 2.142414    & 18.941983  & 7.690220    & 11.251763   & 112.244910  & 43.849197   & 68.395713    \\
				}\mydata

				\addplot[mark=*] table [x=nthread, y expr=\thisrow{t8000phase1}/\thisrow{t8000}] {\mydata};
				\addplot[dashed,mark=square,mark options={solid},blue] table [x=nthread, y expr=\thisrow{t8000phase2}/\thisrow{t8000}] {\mydata};

				\legend{phase 1\\phase 2\\}

			\end{semilogxaxis}
		\end{tikzpicture}
	}
	\hfill
	\subcaptionbox{Parallel speedup of ParaHT}[.49\textwidth]{%
		\begin{tikzpicture}
			\begin{loglogaxis}[
					xlabel=Number of threads,
					ylabel=Parallel speedup,
					legend style={at={(0.03,0.97)},anchor=north west,nodes={scale=0.9, transform shape}},
					ytick={1,2,4,8,10},
					yticklabels={1,2,4,8,10},
					xtick={1,2,4,8,14,28},
					xticklabels={1,2,4,8,14,28},
					grid=major,
					width=0.5\textwidth,
				]

				\pgfplotstableread[row sep=\\,col sep=&]{
					nthread & t1000    & t1000phase1 & t1000phase2 & t2000     & t2000phase1 & t2000phase2 & t4000      & t4000phase1 & t4000phase2 & t8000       & t8000phase1 & t8000phase2  \\
					1       & 1.678905 & 0.621871    & 1.057035    & 14.948054 & 5.650263    & 9.297791    & 132.776269 & 46.639628   & 86.136641   & 1183.129275 & 386.106099  & 797.023176  \\
					2       & 1.017018 & 0.415308    & 0.601710    & 7.863060  & 2.834056    & 5.029004    & 73.398768  & 26.542786   & 46.855981   & 744.187196  & 249.537004  & 494.650192  \\
					4       & 0.845592 & 0.357868    & 0.487725    & 4.816636  & 2.140299    & 2.676337    & 40.209688  & 15.154549   & 25.055139   & 403.227167  & 134.187146  & 269.040021  \\
					8       & 0.795709 & 0.322723    & 0.472986    & 3.723799  & 1.593960    & 2.129839    & 22.943964  & 10.004285   & 12.939679   & 186.901355  & 69.896575   & 117.004780  \\
					14      & 0.750277 & 0.298037    & 0.452240    & 3.571014  & 1.447620    & 2.123394    & 18.832407  & 7.951350    & 10.881056   & 126.416005  & 51.091414   & 75.324591    \\
					28      & 0.751109 & 0.300108    & 0.451001    & 3.524856  & 1.382442    & 2.142414    & 18.941983  & 7.690220    & 11.251763   & 112.244910  & 43.849197   & 68.395713    \\
				}\mydata

				\addplot[mark=*] table [x=nthread, y expr=1.678905/\thisrow{t1000}] {\mydata};
				\addplot[dashed,mark=*,mark options={solid},blue] table [x=nthread, y expr=14.948054/\thisrow{t2000}] {\mydata};
				\addplot[mark=square,red] table [x=nthread, y expr=132.776269/\thisrow{t4000}] {\mydata};
				\addplot[dashed,mark=square,mark options={solid},purple] table [x=nthread, y expr=1183.129275/\thisrow{t8000}] {\mydata};

				\legend{1000\\2000\\4000\\8000\\}

			\end{loglogaxis}
		\end{tikzpicture}
	}
	\hfill
	\subcaptionbox{Parallel speedup of phase 1}[.49\textwidth]{%
		\begin{tikzpicture}
			\begin{loglogaxis}[
					xlabel=Number of threads,
					ylabel=Parallel speedup,
					legend style={at={(0.03,0.97)},anchor=north west,nodes={scale=0.9, transform shape}},
					ytick={1,2,4,8,10},
					yticklabels={1,2,4,8,10},
					xtick={1,2,4,8,14,28},
					xticklabels={1,2,4,8,14,28},
					grid=major,
					width=0.5\textwidth,
				]

				\pgfplotstableread[row sep=\\,col sep=&]{
					nthread & t1000    & t1000phase1 & t1000phase2 & t2000     & t2000phase1 & t2000phase2 & t4000      & t4000phase1 & t4000phase2 & t8000       & t8000phase1 & t8000phase2  \\
					1       & 1.678905 & 0.621871    & 1.057035    & 14.948054 & 5.650263    & 9.297791    & 132.776269 & 46.639628   & 86.136641   & 1183.129275 & 386.106099  & 797.023176  \\
					2       & 1.017018 & 0.415308    & 0.601710    & 7.863060  & 2.834056    & 5.029004    & 73.398768  & 26.542786   & 46.855981   & 744.187196  & 249.537004  & 494.650192  \\
					4       & 0.845592 & 0.357868    & 0.487725    & 4.816636  & 2.140299    & 2.676337    & 40.209688  & 15.154549   & 25.055139   & 403.227167  & 134.187146  & 269.040021  \\
					8       & 0.795709 & 0.322723    & 0.472986    & 3.723799  & 1.593960    & 2.129839    & 22.943964  & 10.004285   & 12.939679   & 186.901355  & 69.896575   & 117.004780  \\
					14      & 0.750277 & 0.298037    & 0.452240    & 3.571014  & 1.447620    & 2.123394    & 18.832407  & 7.951350    & 10.881056   & 126.416005  & 51.091414   & 75.324591    \\
					28      & 0.751109 & 0.300108    & 0.451001    & 3.524856  & 1.382442    & 2.142414    & 18.941983  & 7.690220    & 11.251763   & 112.244910  & 43.849197   & 68.395713    \\
				}\mydata

				\addplot[mark=*,black] table [x=nthread, y expr=0.621871/\thisrow{t1000phase1}] {\mydata};
				\addplot[dashed,mark=*,mark options={solid},blue] table [x=nthread, y expr=5.650263/\thisrow{t2000phase1}] {\mydata};
				\addplot[mark=square,mark options={solid},red] table [x=nthread, y expr=46.639628/\thisrow{t4000phase1}] {\mydata};
				\addplot[dashed,mark=square,mark options={solid},purple] table [x=nthread, y expr=386.106099/\thisrow{t8000phase1}] {\mydata};

				\legend{1000\\2000\\4000\\8000\\}

			\end{loglogaxis}
		\end{tikzpicture}
	}
	\hfill
	\subcaptionbox{Parallel speedup of phase 2}[.49\textwidth]{%
		\begin{tikzpicture}
			\begin{loglogaxis}[
					xlabel=Number of threads,
					ylabel=Parallel speedup,
					legend style={at={(0.03,0.97)},anchor=north west,nodes={scale=0.9, transform shape}},
					ytick={1,2,4,8,10},
					yticklabels={1,2,4,8,10},
					xtick={1,2,4,8,14,28},
					xticklabels={1,2,4,8,14,28},
					grid=major,
					width=0.5\textwidth,
				]

				\pgfplotstableread[row sep=\\,col sep=&]{
					nthread & t1000    & t1000phase1 & t1000phase2 & t2000     & t2000phase1 & t2000phase2 & t4000      & t4000phase1 & t4000phase2 & t8000       & t8000phase1 & t8000phase2  \\
					1       & 1.678905 & 0.621871    & 1.057035    & 14.948054 & 5.650263    & 9.297791    & 132.776269 & 46.639628   & 86.136641   & 1183.129275 & 386.106099  & 797.023176  \\
					2       & 1.017018 & 0.415308    & 0.601710    & 7.863060  & 2.834056    & 5.029004    & 73.398768  & 26.542786   & 46.855981   & 744.187196  & 249.537004  & 494.650192  \\
					4       & 0.845592 & 0.357868    & 0.487725    & 4.816636  & 2.140299    & 2.676337    & 40.209688  & 15.154549   & 25.055139   & 403.227167  & 134.187146  & 269.040021  \\
					8       & 0.795709 & 0.322723    & 0.472986    & 3.723799  & 1.593960    & 2.129839    & 22.943964  & 10.004285   & 12.939679   & 186.901355  & 69.896575   & 117.004780  \\
					14      & 0.750277 & 0.298037    & 0.452240    & 3.571014  & 1.447620    & 2.123394    & 18.832407  & 7.951350    & 10.881056   & 126.416005  & 51.091414   & 75.324591    \\
					28      & 0.751109 & 0.300108    & 0.451001    & 3.524856  & 1.382442    & 2.142414    & 18.941983  & 7.690220    & 11.251763   & 112.244910  & 43.849197   & 68.395713    \\
				}\mydata

				\addplot[mark=*,black] table [x=nthread, y expr=1.057035/\thisrow{t1000phase2}] {\mydata};
				\addplot[dashed,mark=*,mark options={solid},blue] table [x=nthread, y expr=9.297791/\thisrow{t2000phase2}] {\mydata};
				\addplot[mark=square,red] table [x=nthread, y expr=86.136641/\thisrow{t4000phase2}] {\mydata};
				\addplot[dashed,mark=square,mark options={solid},purple] table [x=nthread, y expr=797.023176/\thisrow{t8000phase2}] {\mydata};

				\legend{1000\\2000\\4000\\8000\\}

			\end{loglogaxis}
		\end{tikzpicture}
	}
	\caption{Parallel speedup and relative runtime of our algorithm and its phases.}
	\label{fig:phase1phase2comparison}
\end{figure}

\paragraph*{Tests on saddle point problems}

Following Bujanovic et al.\ \cite{bujanovic2018householder}, we also test our algorithm on saddle point problems. These are pencils of the form:
\begin{equation*}
	(A, B) = \left(\begin{bmatrix}
		X & Y \\ Y^* & 0
	\end{bmatrix},
	\begin{bmatrix}
		I & 0 \\ 0 & 0
	\end{bmatrix}\right),
\end{equation*}
where $I$ is the identity matrix, $Y$ is a random matrix and $X$ is a random symmetric positive definite matrix. By choosing the dimension of $X$ and the identity matrix, we can control the number of infinite eigenvalues in the pencil. In our experiments, the dimensions have been chosen so that $25\%$ of the eigenvalues of the pencil are infinite. Such a pencil is particularly difficult for HouseHT and IterHT.

We note that while pencils with a large number of infinite eigenvalues exist in practice, it is often possible to deflate these eigenvalues prior to the Hessenberg-triangular reduction. LAPACK can detect infinite eigenvalues based on the sparsity structure of the pencil and a different deflation technique we presented in a previous paper \cite{steel2023} can deflate eigenvalues that are not exactly infinite; but can be considered numerically infinite in finite precision.

\begin{figure}[htb]
	\centering
	\begin{tikzpicture}
		\begin{loglogaxis}[
				xlabel=$n$,
				ylabel=Speedup,
				legend style={at={(0.97,0.03)},anchor=south east,nodes={scale=0.9, transform shape}},
				ytick={0.5,1,2,4,8},
				yticklabels={0.5,1,2,4,8},
				xtick={500,1000,2000,4000,8000,16000},
				xticklabels={500,1000,2000,4000,8000,16000},
				ymax=12,
				grid=major,
				height=0.4\textwidth,
				width=0.8\textwidth,
			]

			\pgfplotstableread[row sep=\\,col sep=&]{
				n       & tlapack       & thouseht      & tmdk\\
				500     &     0.147663  &     1.199500  &     0.211133\\
				707     &     0.367131  &     2.569658  &     0.397448\\
				1000    &     0.959683  &     4.832383  &     0.760001\\
				1414    &     2.496546  &    16.597635  &     1.517605\\
				2000    &     7.532678  &    38.000038  &     3.504881 \\
				2828    &    20.921243  &    65.300489  &     8.013165  \\
				4000    &    57.721039  &   126.832965  &    18.785760   \\
				5657    &   168.275135  &   276.970838  &    45.696481   \\
				8000    &   474.415290  &   636.032131  &   114.442442    \\
				11314   &  1390.255426  &  1392.270400  &   345.085465    \\
				16000   &  4128.208167  &  3545.268952  &  1067.007897    \\
			}\mydata

			\addplot[mark=*,black] table [x=n, y expr=\thisrow{tlapack}/\thisrow{tmdk}] {\mydata};
			\addplot[mark=square,blue] table [x=n, y expr=\thisrow{thouseht}/\thisrow{tmdk}] {\mydata};

			\legend{LAPACK\\HouseHT\\}

		\end{loglogaxis}
	\end{tikzpicture}
	\caption{The speedup ParaHT achieves over other algorithms for saddle point pencils of varying size. IterHT is not listed because it failed to converge within 10 iterations of iterative refinement.}
	\label{fig:saddlepointtest}
\end{figure}
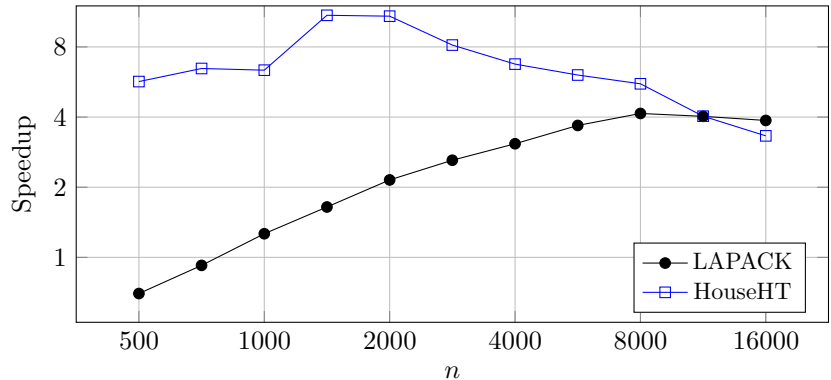

Figure~\ref{fig:saddlepointtest} shows the speedup of ParaHT for saddle point problems of varying sizes. As expected, the speedup over LAPACK has not changed significantly, because the runtimes of LAPACK and ParaHT do not depend on the number of infinite eigenvalues. The speedup over HouseHT is much larger than before; because HouseHT must do more iterative refinement to deal with the ill-conditioned matrix. IterHT is even more sensitive to the presence of a large number of infinite eigenvalues and fails to converge.

\section{Conclusion}

We have presented a parallel algorithm for\\ Hessenberg-triangular reduction. The algorithm consists of two stages: a reduction to r-Hessenberg-triangular form and a reduction to Hessenberg-triangular form. Two-stage reductions typically require more operations than one-stage reductions but are easier to parallelize and our new formulation is no different. Experiments in a shared memory environment have shown that our algorithm is accurate and can outperform state-of-the-art sequential algorithms using parallel BLAS. Future work will need to determine whether our algorithm can also outperform parallelized versions of these algorithms. As an extra advantage, our algorithm does not suffer a performance penalty when infinite eigenvalues are present in the pencil. 

\section*{Acknowledgements}

We thank Bujanovic, Karlsson, and Kressner for providing us with their implementation of HouseHT to compare against.

\clearpage

\bibliographystyle{plain}
\bibliography{references}

\clearpage
\section*{Appendix}

\begin{algorithm}[htb]
   \caption[Blocked stage 1]{Blocked reduction to r-Hessenberg-triangular form \cite{dackland1998scalapack}}
	\label{alg:blockedstage1}
	\begin{algorithmic}[1]
		\REQUIRE $n \times n$ pencil $(A,B)$, with $B$ upper triangular
		\ENSURE decomposition $(A_{orig},B_{orig}) = Q(A,B)Z^*$, with $(A,B)$ in r-Hessenberg-triangular form

      \STATE{$Q = I$}
      \STATE{$Z = I$}
		\FOR{$j = 1:n_b:n-2$}
			\STATE{$j_1 = j$}
			\STATE{$j_2 = \min(n,j+n_b-1)$}
      	\STATE{$n_{blocks} = \lceil(n-nb-j+1)/((p-1)n_b)\rceil$}
			\FOR{$k = n_{blocks}-1:-1:0$}\label{alg:blockedstage1:leftmul}
				\STATE{$i_1 = j+n_b + k(p-1)n_b$}
				\STATE{$i_2 = \min(n,i_1+p*n_b-1)$}
				\STATE{$A(i_1:i_2,j_1:j_2) = \hat{Q}_k^j\hat{R}$}
				\COMMENT{QR factorization}
				\STATE{$A(i_1:i_2,j_1:j_2) = \hat{R}$}
				\STATE{$A(i_1:i_2,j_2+1:n) = \hat{Q}_k^{j*}A(i_1:i_2,j_2+1:n)$}\label{line:stage1-1}
				\STATE{$B(i_1:i_2,i_1:n) = \hat{Q}_k^{j*}B(i_1:i_2,i_1:n)$}\label{line:stage1-2}
				\STATE{$Q(1:n,i_1:i_2) = Q(1:n,i_1:i_2)\hat{Q}_k^{j}$}\label{line:stage1-3}
			\ENDFOR\label{alg:blockedstage1:leftmul2}

			\FOR{$i = j+n_b + (n_{blocks}-1)(p-1)n_b:-(p-1)n_b:j+n_b$}\label{alg:blockedstage1:rightmul}
				\STATE{$i_1 = i$}
				\STATE{$i_2 = \min(n,i+p*n_b-1)$}

				\STATE{$B(i_1:i_2,i_1:i_2) = \tilde{R}\tilde{Q}$}\COMMENT{RQ factorization}
				\STATE{$\tilde{Q}(1:\min(n_b,i_2-i_1+1),1:i_2-i_1+1) = L\hat{Z}$}\COMMENT{LQ factorization}
				\STATE{$A(1:n,i_1:i_2) = A(1:n,i_1:i_2)\hat{Z}$}\label{line:stage1-4}
				\STATE{$B(1:i_2,i_1:i_2) = B(1:i_2,i_1:i_2)\hat{Z}$}\label{line:stage1-5}
				\STATE{$Z(1:n,i_1:i_2) = Z(1:n,i_1:i_2)\hat{Z}$}\label{line:stage1-6}
			\ENDFOR\label{alg:blockedstage1:rightmul2}
		\ENDFOR
	\end{algorithmic}
\end{algorithm}

\begin{algorithm}[htb]
   \caption[Unblocked stage 2]{Reduction to Hessenberg-triangular form without blocking}
	\label{alg:unblockedstage2}
	\begin{algorithmic}[1]
		\REQUIRE $n \times n$ pencil $(A,B)$ in r-Hessenberg-triangular form
		\ENSURE decomposition $(A_{orig},B_{orig}) = Q(A,B)Z^*$, with $(A,B)$ in Hessenberg-triangular form

      \STATE{$Q = I$}
      \STATE{$Z = I$}
		\FOR{$j = 1:n-2$}
			\STATE{$n_{blocks} = 1 + \lfloor \frac{n-j-2}{r} \rfloor$}
			\FOR{$k = 0:n_{blocks}-1$}
				\STATE{$j_b = j + \max(0,(k-1)r + 1) $}
				\STATE{$i_1 = j+kr+1$}
				\STATE{$i_2 = \min(j + (k+1)r,n)$}
				\STATE{$i_3 = \min(j + (k+2)r,n)$}
				\STATE{Generate a reflector $\hat{Q}_k^j = I - \tau_l v_lv_l^*$ that reduces $A(i_1:i_2,j_b)$}
				\STATE{$A(i_1:i_2,j_b:n) = \hat{Q}_k^jA(i_1:i_2,j_b:n)$}
				\STATE{$B(i_1:i_2,i_1:n) = \hat{Q}_k^jB(i_1:i_2,i_1:n)$}
				\STATE{$Q(:,i_1:i_2) = Q(:,i_1:i_2)\hat{Q}_k^j$}
				\STATE{$B(i_1:i_2,i_1:i_2) = \tilde{R}\tilde{Q}$}\COMMENT{RQ factorization}
				\STATE{Generate a reflector $\hat{Z}_k^j = I - \tau_r v_rv_r^*$ that reduces $\tilde{Q}(1,1:i_2-i_1+1)$}
				\STATE{$A(1:i_3,i_1:i_2) = A(1:i_3,i_1:i_2)\hat{Z}_k^j$}
				\STATE{$B(1:i_2,i_1:i_2) = B(1:i_2,i_1:i_2)\hat{Z}_k^j$}
				\STATE{$Z(:,i_1:i_2) = Z(:,i_1:i_2)\hat{Z}_k^j$}
			\ENDFOR
		\ENDFOR
	\end{algorithmic}
\end{algorithm}

\begin{algorithm}
   \caption[Generate stage of stage 2]{Generate phase of stage two}
	\label{alg:stage2generate}
	\begin{algorithmic}[1]
		\REQUIRE $n \times n$ pencil $(A,B)$ in r-Hessenberg-triangular form, where $A$ is in Hessenberg form in columns $1:j_1-1$.
		\FOR{$j = j_1:j_1+q-1$}
			\STATE{$n_{blocks} = 2 + \lfloor \frac{n-j-1}{r} \rfloor$}
			\FOR{$k = 0:n_{blocks}-1$}
				\STATE{$j_b = j + \max(0,(k-1)r + 1) $}
				\STATE{$i_1 = j+kr+1$}
				\STATE{$i_2 = \min(j + (k+1)r,n)$}
				\STATE{$i_3 = \min(j + (k+2)r,n)$}
				\STATE{$i_4 = j_1+1+\max(0,(k+j-j_1-q+2)r)$}

				\FOR{$\hat{j} = j_1:j-1$}
					\STATE{$\hat{i_1} = \hat{j} + kr+1$}
					\STATE{$\hat{i_2} = \min(\hat{j} + (k+1)r,n)$}
					\IF{$\hat{i_2} - \hat{i_1} \ge 1$}
						\STATE{$A(\hat{i_1}:\hat{i_2},j_b) = \hat{Q}_k^{\hat{j}}A(\hat{i_1}:\hat{i_2},j_b)$}
						\IF{$i_1 + r - 1 \le n$}
							\STATE{$B(\hat{i_1}:\hat{i_2},i_1+r-1) = \hat{Q}_k^{\hat{j}}B(\hat{i_1}:\hat{i_2},i_1+r-1)$}
						\ENDIF
					\ENDIF
				\ENDFOR
				\STATE{Generate a reflector $\hat{Q}_k^j = I - \tau_l v_lv_l^*$ that reduces $A(i_1:i_2,j_b)$}
				\STATE{$A(i_1:i_2,j_b) = \hat{Q}_k^jA(i_1:i_2,j_b)$}
				\STATE{$B(i_1:i_2,i_1:i_2) = \hat{Q}_k^jB(i_1:i_2,i_1:i_2)$}
				\STATE{$B(i_1:i_2,i_1:i_2) = \tilde{R}\tilde{Q}$}\COMMENT{RQ factorization}
				\STATE{Generate a reflector $\hat{Z}_k^j = I - \tau_r v_rv_r^*$ that reduces $\tilde{Q}(1,1:i_2-i_1+1)$}
				\STATE{$A(i_4:i_3,i_1:i_2) = A(i_4:i_3,i_1:i_2)\hat{Z}_k^j$}
				\STATE{$B(i_4:i_2,i_1:i_2) = B(i_4:i_2,i_1:i_2)\hat{Z}_k^j$}
			\ENDFOR
		\ENDFOR
	\end{algorithmic}
\end{algorithm}

\begin{algorithm}
   \caption[Application stage of stage 2]{Application phase of stage two}
	\label{alg:stage2application}
	\begin{algorithmic}[1]
		\REQUIRE $n \times n$ pencil $(A,B)$ in r-Hessenberg-triangular form, where $A$ is in Hessenberg form in columns $1:j_1-1$.
		\STATE{$n_{blocks} = 1 + \lfloor \frac{n-j_1-2}{r} \rfloor$}
		\FOR{$k = n_{blocks}-1:-1:0$}
			\STATE{$i_5 = j_1+1+\max(0,(k-q+2)r)$}
			\FOR{$j = j_1+1:j_1+q-1$}
				\STATE{$i_1 = j+kr+1$}
				\STATE{$i_2 = \min(j + (k+1)r,n)$}
				\STATE{$i_4 = j_1+\max(0,(k+j-j_1-q+2)r)$}
				\STATE{$A(i_5:i_4,i_1:i_2) = A(i_5:i_4,i_1:i_2)\hat{Z}_k^j$}
				\STATE{$B(i_5:i_4,i_1:i_2) = B(i_5:i_4,i_1:i_2)\hat{Z}_k^j$}
			\ENDFOR
			\STATE{$\hat{Z}_k = \hat{Z}_k^{j_1}\hat{Z}_k^{j_2} \dots \hat{Z}_k^{j_1+q-1} $}\COMMENT{Form compact WY}
			\STATE{$i_1 = j_1+kr+1$}
			\STATE{$i_2 = \min(j_1+q-1+(k+1)r,n)$}
			\STATE{$i_5 = j_1+\max(0,(k-q+2)r)$}
			\STATE{$A(1:i_5,i_1:i_2) = A(1:i_5,i_1:i_2)\hat{Z}_k$}
			\STATE{$B(1:i_5,i_1:i_2) = B(1:i_5,i_1:i_2)\hat{Z}_k$}
			\STATE{$Z(:,i_1:i_2) = Z(:,i_1:i_2)\hat{Z}_k$}
		\ENDFOR
		\FOR{$k = n_{blocks}-1:-1:0$}
			\STATE{$\hat{Q}_k = \hat{Q}_k^{j_1}\hat{Q}_k^{j_2} \dots \hat{Q}_k^{j_1+q-1} $}\COMMENT{Form compact WY}
			\STATE{$i_1 = j_1+kr+1$}
			\STATE{$i_2 = \min(j_1+q-1+(k+1)r,n)$}
			\STATE{$i_5 = j_1+q-1 + \max(0,(k-1)r + 1) $}
			\STATE{$i_6 = j_1+q+(k+1)r-1 $}
			\STATE{$A(i_1:i_2,i_5:n) = \hat{Q}_k^*A(i_1:i_2,i_5:n)$}
			\STATE{$B(i_1:i_2,i_6:n) = \hat{Q}_k^*B(i_1:i_2,i_6:n)$}
			\STATE{$Q(:,i_1:i_2) = Q(:,i_1:i_2)\hat{Q}_k$}
		\ENDFOR
	\end{algorithmic}
\end{algorithm}
\end{document}